\documentclass[useAMS]{mn2e}
\usepackage{psfig}

\usepackage{times}
\def\spose#1{\hbox to 0pt{#1\hss}}
\newcommand\lsim{\mathrel{\spose{\lower 3pt\hbox{$\mathchar"218$}}
     \raise 2.0pt\hbox{$\mathchar"13C$}}}
\newcommand\gsim{\mathrel{\spose{\lower 3pt\hbox{$\mathchar"218$}}
     \raise 2.0pt\hbox{$\mathchar"13E$}}}
\def\ltsima{$\; \buildrel < \over \sim \;$}
\def\lsim{\lower.5ex\hbox{\ltsima}}
\def\gtsima{$\; \buildrel > \over \sim \;$}
\def\gsim{\lower.5ex\hbox{\gtsima}}

\title[The $\gamma$--ray brightest days of the blazar 3C 454.3] 
{The $\gamma$--ray brightest days of the blazar 3C 454.3}

\author[Bonnoli et al.] 
{G. Bonnoli,\thanks{E--mail:
giacomo.bonnoli@brera.inaf.it} 
G. Ghisellini, L. Foschini,
 F. Tavecchio and G. Ghirlanda\\
INAF -- Osservatorio Astronomico di Brera, via E. Bianchi 46, I--23807 Merate, Italy \\
}

\begin{document}

% \date{Accepted 1988 December 15. Received 1988 December 14; 
% in original form 1988 October 11}

\pagerange{\pageref{firstpage}--\pageref{lastpage}} \pubyear{2007}

\maketitle

\label{firstpage}

\begin{abstract}
In the first week of December 2009, the blazar 3C 454.3
became the brightest high energy source in the sky.
Its photon flux reached and surpassed the level of
$10^{-5}$ ph cm$^{-2}$ s$^{-1}$ above 100 MeV.
The {\it Swift} satellite observed the source several times
during the period of high $\gamma$--ray flux, and we can construct
really simultaneous spectral energy distributions (SED) before, during,
and after the luminosity peak.
Our main findings are:
i) the optical, X--ray and $\gamma$--ray fluxes correlate;
ii) the $\gamma$--ray flux varies quadratically (or even more)
with the optical flux;
iii) a simple one--zone synchrotron inverse Compton model can account
for all
the considered SED;
iv) in this framework the $\gamma$--ray vs optical flux correlation
can be explained if the magnetic field is slightly fainter when the overall
jet luminosity is stronger;
v) the power that the jet spent to produce the peak $\gamma$--ray luminosity
is of the same order, or larger, than the accretion disk luminosity.
%During the flare, the total jet power surely surpassed the accretion power. 
\end{abstract}
\begin{keywords}
galaxies: active--galaxies: jets--galaxies: individual: 3C454.3 ---
radiation mechanisms: non--thermal
\end{keywords}

\section{Introduction}

Among the Flat--Spectrum Radio Quasar (FSRQ) class of blazars, 
3C 454.3 ($z=0.859$, Jackson \& Browne 1991)
is one of the brightest and most variable sources.
In  April--May 2005 it underwent a dramatic optical outburst, reaching
its historical maximum with $R=12.0$ (Villata et al. 2006).
This exceptional event triggered observations at X--ray energies with the \emph{INTEGRAL}\ 
(Pian et al. 2006) and the {\it Swift} satellites (Giommi et al. 2006).
These detected the source up to 200 ~keV and in a brightened 
high energy state.
This strong optical and hard X--ray activity was followed by a radio outburst 
with about one year delay (Villata et al. 2007).
Despite this extraordinary activity, the lack of a $\gamma$--ray satellite
at that time prevented us from concluding that this activity was due to a real
increase of the jet power, since the integrated optical luminosity was of the
same order of the $\gamma$--ray luminosity observed, years before, by the 
EGRET instrument onboard the {\it Compton Gamma Ray Observatory (CGRO)} satellite
(Nandikotkur et al. 2007).
After the launch of {\it AGILE}, 3C 454.3 was seen to be very bright
(Vercellone et al. 2007) and variable in the $\gamma$--ray band, 
being often the brightest $\gamma$--ray blazar in the sky (Vercellone et al. 2010),
even if the simultaneous optical states were much 
fainter than during the 2005 flare.
Many other active phases followed also in summer 2008  just after the launch
of the \textit{Fermi} satellite (Tosti et al. 2008;
Abdo et al. 2009a), and in summer--fall 2009, when the source was detected in active 
state all across the entire electromagnetic spectrum leading to many 
Astronomer's Telegrams (ATels) (Gurwell 2009; Buxton et al. 2009;
Hill 2009; Striani et al. 2009a; Bonning et al. 2009;  Villata et al. 2009).
This activity climaxed in December 2009, when an extraordinary
$\gamma$--ray flare occurred on December the 2nd, seen by AGILE (Striani et
al. 2009b), \textit{Fermi}/LAT (Escande \& Tanaka 2009),
{\it Swift}/XRT (Sakamoto et al. 2009), {\it Swift}/BAT (Krimm et al. 2009). 

In the case of the 2007 flare it was possible to model its spectral energy
distribution (SED) in different states with a jet whose total power was 
nearly constant in time 
(Katarzynski \& Ghisellini 2007), but whose dissipation region was 
located at different distances $R_{\rm diss}$ from the central power--house 
(Ghisellini et al. 2007).
In this model, the magnetic field energy density scales as 
$U_B\propto (\Gamma R_{\rm diss})^{-2}$, where $\Gamma$ is the bulk Lorentz
factor. % while 
 Conversely, the radiation energy density 
$U_{\rm ext}$ produced by the broad line clouds scales as $\Gamma^2$ 
and is independent of $R_{\rm diss}$ as long as the dissipation region 
is within the broad line region (BLR).
Even if the latter is most likely stratified, the contribution of the Ly--$\alpha$
line to the BLR luminosity is dominant, and we identify the size of the 
BLR with the shell mostly emitting the Ly--$\alpha$ photons.
As a consequence, $U_B/U_{\rm ext}\sim L_{\rm opt}/L_\gamma 
\propto R_{\rm diss}^{-2}\Gamma^{-4}$.   
Even with a constant and dominating $L_\gamma$ and a constant $\Gamma$,
we can have large variations of the optical flux by changing $R_{\rm diss}$.
These ideas were borne out in Ghisellini et al. (2007), immediately after the 
{\it AGILE} detection of a very bright $\gamma$--ray state accompanied by a 
moderate optical flux: the optical synchrotron flux can vary by a large 
amount even if the jet continues to carry the same power and 
always dissipates the same fraction of it. 
This assumes and implies, however, that the bulk of the electromagnetic output
is in the high energy (inverse Compton) component, that in this case should remain
quasi--steady.

Instead the large outburst seen in December 2009 witnessed a dramatic increase
of this component. 
In this case it is very likely that the power dissipated by the jet indeed increased,
and it is likely that this is the consequence of an increase of total power carried
by the jet.
The luminosity reached in $\gamma$--rays exceeded $L_\gamma\sim 3\times 10^{49}$ 
erg s$^{-1}$, a factor $\sim$30 larger than the EGRET luminosity, 
and also larger than the 2007 {\it AGILE} state.
The goal of our present study is first to characterise the emission properties of 
3C 454.3 during the November--December 2009 period, then to find the physical properties
of the jet during the flare, and finally to compare those with the luminosity
of the accretion disk.
In fact, as we shall see, the power of the jet of 3C 454.3 in the first week 
of December 2009 likely surpassed the luminosity of the accretion disk.

We use $H_0=70$ km s$^{-1}$ Mpc$^{-1}$  and $\Omega_{\rm M}=0.3$, 
$\Omega_{\Lambda}=0.7$ and use the notation $Q=10^x Q_x$ in cgs units.

\section{Data \& analysis}
\label{analysis}

\subsection{Data sets}

This study is focused on the analysis of the multi--wavelength (MWL) data gathered by the
\emph{Fermi} and \emph{Swift} satellites in the period between November 1 and
December 14, 2009. 
 From these data we obtained day--scale resolved light curves along the whole
time span, in the  $\gamma$--ray, X--ray and optical--UV band.
Moreover, for some selected days we derived the spectral information needed
to study the simultaneous MWL SED. 
These data sets are labelled according to the observation date.

In order to fully characterise the evolution of the SED across the whole
dynamic range of $\gamma$--ray flux observed by \emph{Fermi}, we also built 
a ``\emph{low}'' SED, deriving a high--energy (HE) spectrum of
3C 454.3 from an earlier period of observation when the source was significantly
weaker in $\gamma$--rays, and matching it with \textit{Swift} data from the same period.

Finally, in order to better investigate the UV portion of the SED, we matched
an archival UV spectrum of 3C454.3 observed by the \emph{GALEX} satellite 
with contemporary optical--UV data taken by \emph{Swift}.

\subsection{{\it Fermi}/LAT}

We retrieved from the NASA 
database\footnote{http://fermi.gsfc.nasa.gov/} the publicly available data 
from the LAT $\gamma$-ray telescope
onboard of the \emph{Fermi} satellite (Atwood et al. 2009). 
We selected the good quality ("DIFFUSE" class) events observed within
10$^{\circ}$ from the source position (J2000) ($\alpha$, $\delta$)=(22h53m57.7s, +16$^{\circ}$08'53.56''), 
but  excluding those observed with zenith distance of the arrival direction greater than 105$^{\circ}$,
in order to avoid contamination from  the Earth albedo.
We performed the analysis by means of the standard Science Tools, v. 9.15.2,
including Galactic and isotropic extragalactic backgrounds and the P6 V3 DIFFUSE
instrumental response function.
For each data set (and, in case of spectra, for each energy bin) we 
calculated the livetime, exposure map and diffuse response of the instrument. 
Then we applied to the data an unbinned likelihood algorithm
(\texttt{gtlike}), modelling the source spectrum with a power law model, with
the integral flux in the considered energy band and photon index left as free parameters. 
The minimum statistical confidence level accepted for each time (or time and energy, in the
case of spectra) bin is TS = 10 where TS is the test statistic described in
Mattox et al. 1996 (see also Abdo et al. 2009b).
Quoted errors are statistical, at 1 $\sigma$ level. Systematic flux uncertainties  are
within 10\% at 100 MeV, 5\% at 500 MeV and 20\% at 10 GeV (Rando 2009).

\subsubsection{Light curve}

In the upper panel of Fig. \ref{lc}  we report the integral $\gamma$--ray 
flux above 100 MeV, calculated in 1--day
time bins for the  period  from November 1 to December 14, 2009.
The flux level is always above 10$^{-6}$ cm$^{-2}$ s$^{-1}$,
the clear signature of a coherent $\gamma$--ray active phase.
On December 2 the flux reached
the extreme value of (21.8$\pm$ 1.2) $\times$ 10$^{-6}$ cm$^{-2}$ s$^{-1}$,
the strongest non--GRB flux measured by a $\gamma$--ray satellite.
The corresponding observed $\gamma$--ray luminosity is of 
the same order of PKS 1502+106 in outburst (Abdo et al. 2010a).

\subsubsection{Spectra}

The exceptional flux level of 3C 454.3 in late 2009 allowed for the first time
to derive day--scale $\gamma$--ray spectra with good quality: usually an
inconceivable task for space--based instruments due both to their small
collection area and the weak flux of the high--energy sources.
This opens new opportunities for the study of the 
MWL SED of this FSRQ on short time scales, 
hopefully allowing deeper insight into the engine and the
emission mechanism.

In this study we use LAT spectra produced with data from only a few single days,
selected within the period covered also by {\it Swift} observations.
%,between November the 1st and December the 15th. 
We have chosen to construct the SEDs in order to cover the whole
flux range of the campaign.
We therefore selected November 6,
showing the lowest UVOT fluxes of the whole period; 
November 27, a day in the rising phase of the flare, with intermediate 
$\gamma$--ray flux, and the three days (December 1, 2 and 3) around the peak, 
in order to better investigate the evolution of the SED around this extreme phase. 
For the $\gamma$--ray spectrum of November 6 we decided to add
the data from the following day (November 7), as it showed compatible
$\gamma$--ray flux level, to improve the signal to noise ratio 
(S/N) of the spectrum.
Finally, we selected from the 18--months daily light curve (see e.g. Tavecchio
et al. 2010) a longer time--span (15 Ms), 
starting from 3 December 2008
%\footnote{Mission Elapsed Time (MET)=2.5 $\times 10^{8}$ s.}
to 26 May 2009, when the source was at the lowest flux levels ($F_{> \rm 100  \, MeV} \sim 5
\times 10^{-7}$ cm$^{-2}$ s$^{-1}$)  since the beginning of LAT observations.
The derived HE spectrum allowed to explore the whole
available dynamic range of 3C 454.3 $\gamma$--ray emission. 
For each data set, we split the LAT observed band in logarithmically spaced 
energy bins, and applied in each the unbinned likelihood algorithm. 
In addition to the
request that TS $>$ 10, we further requested that the model predicted at least 5
photons from the source; when needed, we merge in a single, wider bin the high energy 
bins that did not fulfil this last request. 
These quality criteria are similar (and somehow more restrictive) to those
 adopted by Abdo et al. (2010b). %Abdo Spettri
For each spectrum, we crosschecked this analysis performed in bins of energy with an unbinned
analysis on the whole 0.1--100 GeV band, where we modelled the source with a
broken power law model, in agreement with Abdo et al. (2010b). 
The results of the two methods were consistent within
errors, 
even if the analysis in bins of energy, that does not exploit the full
information of the event distribution across the whole LAT bandpass, leads to
wider error bars.
We report in Table \ref{lat} the parameters of the model, while the
SED points are plotted in Fig. \ref{sed}.

\subsection{{\it Swift}}

We analysed the data from observations performed between November 1 
and December 14, 2009. We also analysed data from late December 2008,
contemporary to the lowest $\gamma$--ray state of 3C 454.3 seen by
\textit{Fermi}. 
Finally, UVOT data from a pointing in October 2008 were
analysed, as we wanted to compare this observation with a roughly simultaneous
{\it GALEX} archival UV spectrum of the source (see \S \ref{Lyman} ).

The data from the two narrow field instruments onboard \emph{Swift} 
(Gehrels et al. 2004) have been processed and analysed with 
\texttt{HEASoft v. 6.8} and  with the CALDB updated on  December 30, 2009.

\subsubsection{X--Ray Telescope}

Data from the X--ray Telescope (XRT) (0.2--10 keV, Burrows et al. 2005) 
were analysed using the \texttt{xrtpipeline} task,
set for the photon counting or window timing modes and having selected single pixel events (grade $0$). 
In the case of the light curve, we converted the observed counts rates to fluxes
assuming a common $\Gamma = -1.6$ photon index and with absorption fixed at
the  Galactic value 
$N^{\rm Gal}_{\rm H}=6.6\times 10^{20}$~cm$^{-2}$ derived from $A_{B}=0.46$ 
(Raiteri et al. 2008).
The resulting light curve 
%between November 1  and December 14
is shown in the middle panel of Fig. \ref{lc}.

%For deriving the spectra, the output d
We extracted the spectra in order to build SED simultaneous to
\emph{Fermi}/LAT by using default regions.
Data were rebinned in order to have at least 
30 counts per energy bin. 
Power law models have been fitted to the spectra, except for
two cases, the ``\textit{low}'' state and the spectrum from December 2,
when a broken power law model fitted the data significantly better ( ftest $> 99\%$).  
In Tab. \ref{xrt} we report the spectral parameters for the observations used
for modelling the SED.

\subsubsection{UltraViolet--Optical Telescope}

Data from the UV telescope UVOT (Roming et al. 2005) were analysed 
by means of  the \texttt{uvotimsum} and \texttt{uvotsource} 
tasks with a source region of $5''$, 
while the background was extracted from a source--free circular region 
with radius equal to $50''$ (it was not possible to use an annular region, 
because of a nearby source).
% The observed magnitudes are reported in table \ref{uvot}; uncertainties
% include systematics.
A total of 19 UVOT observations have been performed and analyzed between November 1 and 
December 14. 
The extracted fluxes are plotted in the lower panel of Fig. \ref{lc}. 
The Galactic extinction has been corrected assuming
$A_{B}=0.46$ and $R_{V}=A_{V}/E(B-V)= 3.1$ (according to Raiteri et al. 2008), 
and the extinction law of Cardelli et al. (1989). 
The \emph{observed} magnitudes from the 5 pointings used for the MWL SED are reported
in Table \ref{uvot}; uncertainties include systematics. 
The observation made on October 1, 2008, roughly contemporary to the {\it GALEX}
archival spectrum (see \S \ref{galex}) is also reported in the table. 
The UVOT data set used for the modelling of the ``\emph{low}'' state 
is built with pointings made between December 25, 2008 and January 1, 2009. 
%In this case the reported magnitudes and uncertainties are
%respectively the sample mean and sample standard deviation of the magnitudes 
%observed in single observations.
In this case the reported magnitudes are the  mean values over the
observations, while the uncertainties are calculated as $\sigma =
\sqrt{\sigma_{\rm stat}^2 + \sigma_{\rm std}^2}$, where $\sigma_{\rm stat}$ 
is the mean statistical
uncertainty on the single observation and  $\sigma_{\rm std}$ is the sample
standard deviation of the magnitude distribution, and keeps into account the 
variability of the source.

\subsubsection{Burst Alert Telescope}

\textit{Swift} monitors the sky in the hard X--ray band by means of the
Burst Alert Telescope (BAT), a coded mask telescope sensitive in the 15--150
keV band (Barthelmy et al. 2005). 
 We retrieved from the HEASARC
%\footnote{High Energy Astrophysics Science Archive Research Center}
 hard X--ray transient monitor 
database\footnote{http://heasarc.gsfc.nasa.gov/docs/swift/results/transients/}
the daily sampled light curve of 3C 454.3 in the 15 -- 50 keV band, and
converted the count rates to fluxes assuming as reference the BAT count rates
and flux of the Crab Nebula in the same band. 
The corresponding $\nu F_{\nu}$ values are plotted in Fig. \ref{sed}.

\subsection{\emph{GALEX}}
\label{galex}

The Galaxy Evolution Explorer (Martin et al. 2005) is a NASA
satellite, in flight since 28 April 2003 and performing an all--sky
survey in the far UV (FUV, $\sim 154$ nm) and near UV (NUV, $\sim 232$ nm) 
% with $\sim $ 40 cm$^2$ peak response and 
with resolution $R \equiv \lambda / \Delta \lambda \sim$ 200 in NUV slit--less
spectroscopy (Morissey et al. 2007).
We retrieved from the MAST\footnote{Multimission Archive at the Space Telescope
Science Institute, \texttt{http://galex.stsci.edu/GR4/} } 
database a publicly available spectrum of 3C 454.3, obtained from 3 exposures
taken between September 30 and October 2, 2008. 
This spectrum is plotted in Fig. \ref{lyman}, and its relevance is discussed
in \S \ref{Lyman}.  
%Morrissey 2007

% ---------------------------------------------------
\begin{table*} 
\centering
\begin{tabular}{lcccccc}
\hline 
\hline
OBS date& $t_{\rm on}$&$\Gamma_{1}$&$\Gamma_{2}$&$E_{\rm break}$&$F_{ 0.1-100\rm \, GeV}$&$TS$\\
\hline
dd/mm/yy& ks & & &GeV &$10^{-6}\frac{\rm ph}{\rm cm^{2}s}$ & \\
\hline
{\it low}&5.9$\times 10^3$ &2.33$\pm$0.03 &3.37$\pm$0.27 &2.00$\pm$0.38  &0.61$\pm$0.02  &4687  \\
06/11/09       &61               &2.66$\pm$0.21 &3.80$\pm$1.20   &1.10$\pm$0.57 &2.10$\pm$0.30    &218 \\
          		  &                & 2.78$\pm$0.16 &--            &--        &2.15$\pm$0.30 &218 \\
27/11/09       &37               &2.13$\pm$0.10 &3.17$\pm0.54$ &1.94$\pm$0.56 &7.58$\pm$0.68  &900 \\
01/12/09       &44               &2.33$\pm$0.11 &3.15$\pm$0.38 &0.97$\pm$0.34 &10.76$\pm$0.87 &1100 \\
02/12/09       &39               &2.37$\pm$0.06 &3.17$\pm$0.46 &2.70$\pm$0.26 &21.6$\pm$1.2   &2497 \\
03/12/09       &40               &2.28$\pm$0.06 &3.42$\pm$0.87 &5.86$\pm$0.36 &15.94$\pm$0.98 &1915 \\
\hline
\hline
\end{tabular}
\vskip 0.4 true cm
\caption{Results of the $\gamma$--ray analysis on \textit{Fermi}--LAT data; only the
  data set used for the MWL SED are reported. For each data set, the effective
  source on--time $t_{\rm on}$, the photon indices $\Gamma_{1}$ and $\Gamma_{2}$, the
  break energy $E_{\rm break}$, the integral flux in the 0.1 -- 100 GeV band
  $F_{0.1-100\, \rm GeV}$, and the TS value of the unbinned likelihood
  analysis are reported. Uncertainties are statistical only, at 1
  $\sigma$ level; systematics on flux measurement are within 10 \% at 100
  MeV, within 5\% at 500 MeV and within 20\% at 10 GeV.
For the 06/11/009 dataset, the relative weakness of the flux makes perfectly
adequate a simple power law model which is also reported for reference.}
\label{lat}
\end{table*}
% ----------------------------------------------------------

% ---------------------------------------------------
\begin{table*} 
\centering
\begin{tabular}{lccccccccc}
\hline 
\hline   
OBS date &obsID&$t_{\rm exp}$ &$\Gamma_1$          &$\Gamma_2$ &$E_{\rm break}$ &$f_{0}$
  &$F_{\rm 0.2-10\,\rm keV}$ &L$_{\rm 0.2-10\, {\rm  keV}}$ &$\tilde{\chi}^2$ (dof) \\
\hline
dd/mm/yy &  &\rm ks    & &     &keV &10$^{-2} {\rm ph \over cm^2 \, s \, keV}$  
&10$^{-11} { \rm erg \over  cm^2\, s} $ &10$^{47}$ \rm erg/s&  \\ 
\hline  
\emph{low}   & see caption &9.0&1.13$^{+0.30}_{-0.24}$  & 1.13$^{+0.30}_{-0.24}$  
& 1.40$^{+1.00}_{-0.36}$  &  0.15$\pm$0.01     &1.1   &0.34&0.89 (43)   \\ 
06/11/09   &00035050070& 1.0  &1.52$\pm$0.07  &       --       &  --   &  2.2$\pm$ 0.2
&6.6       &2.1&0.96 (38)    \\ 
27/11/09   &00035050073& 0.9   &1.42$\pm0.06$  &       --       &  --   & 3.0 $\pm$ 0.2
& 11.0       &3.1&   1.13 (53) \\ 
01/12/09   &00035050074& 1.1   &1.43$\pm0.04$  &       --       &  --   & 4.5 $\pm$0.3
&16.0&4.5&1.08 (87)    \\ 
02/12/09   &00035050075&1.2   &0.98$^{+0.15}_{-0.19}$  &1.56$\pm$0.07              & 1.2
$\pm$0.2    &   2.2 $\pm$0.2    &17.0        &4.8&0.98 (103)   \\ 
03/12/09   &00035050076&1.0   &1.43$\pm$0.04  &         --     & --    &  5.2 $\pm$0.3
&18.0        &5.2& 0.88 (89)   \\ 
\hline
\hline
\footnotetext{Combination of observations made between 25
  December 2008 and 01 January 2009: obsIDs}
\end{tabular}
\caption{Synopsis of the results of the X--ray analysis performed on
\emph{Swift}/XRT data. Only the data sets used for modelling the MWL
  SED of 3C 454.3 are reported. For each one of these the exposure time $t_{\rm exp}$, the
  photon indices $\Gamma_1$ and $\Gamma_2$, the break energy $E_{\rm break}$, the
  normalisation factor of the differential spectrum at 1 keV $f_0$, the
  observed (not de--absorbed) flux in the 0.2--10 keV band $F_{\rm 0.2-10 \, \rm keV}$, 
  the intrinsic luminosity in the same band $L_{\rm 0.2- 10\, keV}$ 
  and the goodness of fit statistic $\tilde{\chi}^2$ (reduced $\chi^2$) with the
  corresponding number of degrees of freedom (dof) are reported. 
  The ``\emph{low}" data set results from the addition of observations  made
  between 25 Dec 2008 and 01 Jan 2009 (obsID  from 00090023002 up to 00090023008) 
  while the other data sets come from
  single pointings  and are labelled with the date of observation. Spectra have
  been de--reddened according to the $N_{\rm H}= 6.6 \times 10^{20}$ cm$^{-2}$
  value derived from Raiteri et al. (2008). The errors on the observed fluxes are within the 10\%
  level (at 90\% CL).  
}
\label{xrt}

\end{table*}
% ----------------------------------------------------------

% ----------------------------------------------------------
\begin{table*}
\centering
% \begin{tabular}{lccccccc}
\begin{tabular}{lccccccc}
\hline
\hline
OBS date  &obsID &$v$             &$b$             &$u$            &$uvw1$        &$uvm2$        &$uvw2$\\
\hline
01/10/08 (GALEX)  &00031216062&15.42$\pm 0.04$ &15.92$\pm 0.03$ &15.24$\pm 0.04$  &15.41$\pm 0.04$ &15.50$\pm 0.04$ &15.71$\pm 0.04$ \\
\emph{low}  &see caption&16.23$\pm 0.11$ &16.72$\pm 0.07$ &15.93$\pm 0.08$  &15.96$\pm 0.09$ &15.92$\pm 0.08$ &16.18$\pm 0.05$ \\
06/11/09  &00035050070&15.88$\pm 0.06$ &16.33$\pm 0.05$ &15.60$\pm 0.05$  &15.68$\pm 0.05$ &15.68$\pm 0.06$ &15.86$\pm 0.05$ \\
27/11/09  &00035050073&14.77$\pm 0.04$ &15.25$\pm 0.04$ &14.58$\pm 0.04$  &14.85$\pm 0.05$ &14.95$\pm 0.05$ &15.13$\pm 0.04$ \\
01/12/09  &00035050074&14.67$\pm 0.04$ &15.20$\pm 0.03$ &14.57$\pm 0.04$  &14.79$\pm 0.04$ &14.93$\pm 0.05$ &15.16$\pm 0.04$ \\
02/12/09  &00035050075&14.49$\pm 0.04$ &15.02$\pm 0.03$ &14.25$\pm 0.04$  &14.56$\pm 0.04$ &14.67$\pm 0.05$ &14.89$\pm 0.04$ \\
03/12/09  &00035050076&        --      &        --      &       --        &14.58$\pm 0.04$ &14.75$\pm 0.04$ &14.97$\pm 0.04$ \\
\hline
\hline
\end{tabular}
\caption{Summary of \emph{Swift}/UVOT observed magnitudes. 
 Only the data sets that we used for 
 modelling the MWL SED of 3C 454.3 are reported here, out of the total of 19. 
 In addition, we report the UVOT
 observation that we compared to the {\it GALEX} archival spectrum in Fig. \ref{lyman}.
 The data set used for the ``\emph{low}" state includes 7
 UVOT pointings, with obsID starting  from 00090023002 up to 00090023008.
 Magnitudes are not de--reddened; errors include systematics. 
% Fluxes were dereddened adopting $A_B= 0.46$ (Raiteri et al. 2008).
}
\label{uvot}
\end{table*}
% ----------------------------------------------------------

%--------------------------------------------------
\begin{figure}
\vskip -0.5 true cm 
\psfig{file=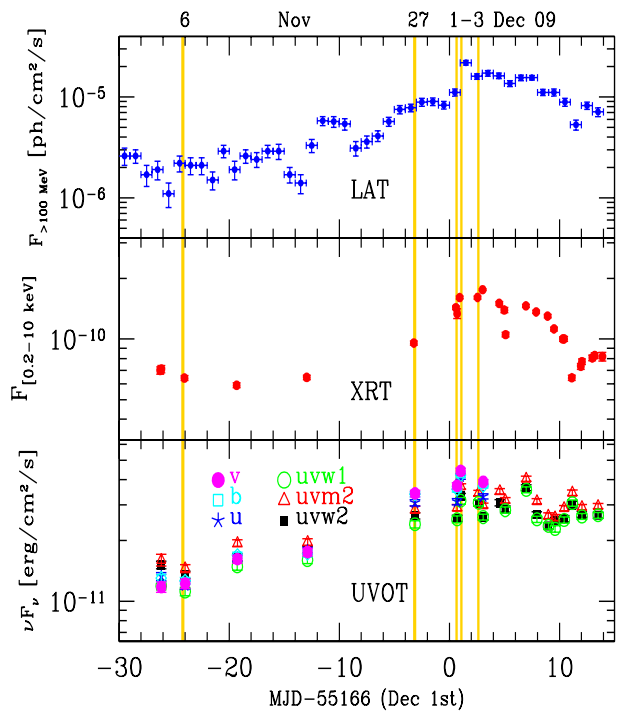,height=11cm,width=9.cm}
\vskip -0.7 true cm
\caption{
Light curve above 100 MeV (top), in the [0.2--10 keV] 
X--ray band (middle panel) and in the different
UVOT filters (bottom panel, as labelled).
Note the logarithmic $y$ axes, and their different scales. Yellow lines mark
the days for which simultaneous multi--wavelenght SED were derived and modelled.
}
\label{lc}
\end{figure}
%--------------------------------------------------
%--------------------------------------------------
\begin{figure}
\vskip -0.1 true cm
\hskip -0.1 true cm
\psfig{file=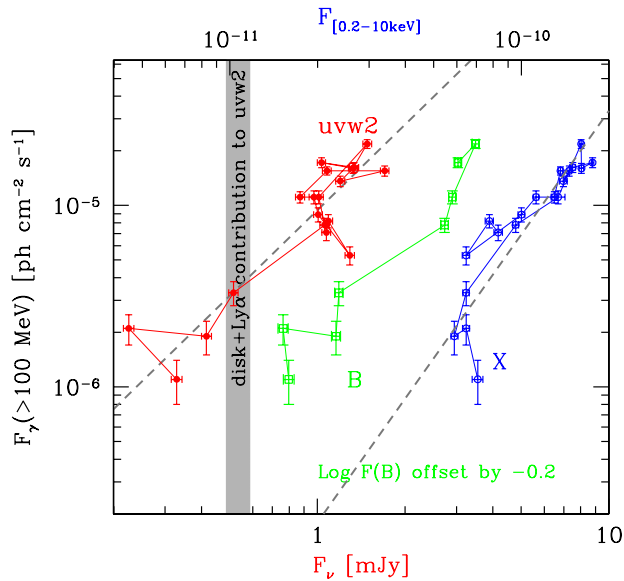,height=9.cm,width=9.cm}
\vskip -0.7 true cm
\caption{
Correlation between the non--thermal UV flux (UVOT filter {\it uvw2}, subtracting the 
disk and the Ly--$\alpha$ contributions), 
non--thermal optical flux (UVOT filter {\it b}, subtracting the disk contribution), 
non--thermal X--rays and the $\gamma$--ray flux.
The grey vertical stripe indicates the contribution of the accretion disk plus
the tail of the Ly--$\alpha$ line entering in the {\it uvw2} frequency range.
The dashed lines are the linear fit to the $\gamma$--ray flux with the {\it uvw2} and X--ray
fluxes (see text). }
\label{uvxg}
\end{figure}
%--------------------------------------------------

%--------------------------------------------------
\begin{figure}
\vskip -0.7 true cm 
\psfig{file=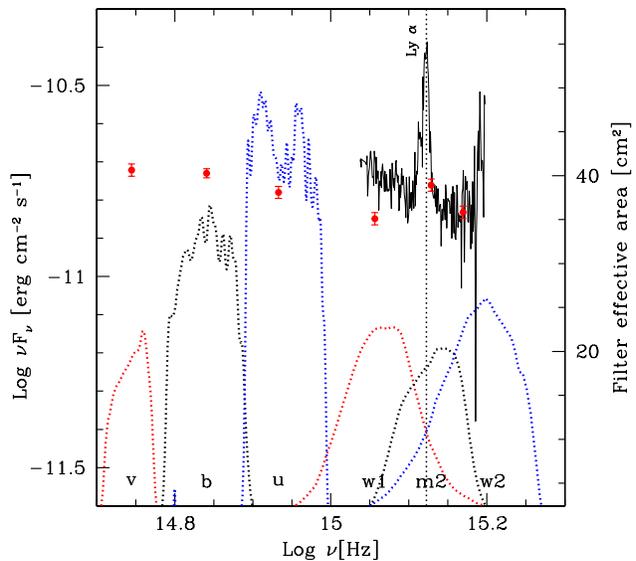,height=8.5cm,width=8.5cm}
\vskip -0.1 true cm
\caption{%Decomposition of the non--thermal continuum, 
%the accretion disk and the contribution from the
%Lyman--$\alpha$ line to the total continuum, and
%the resulting flux in the UVOT filters.
A portion of the UV SED of 3C 454.3 as seen by \emph{GALEX} 
in October 2008 (in black). 
The red SED points are derived from a contemporary UVOT observation. 
At $z=0.859$ the Ly--$\alpha$ line ($\lambda_0=1216$ \AA) is shifted
close to the effective frequency of the \textit{uvm2} filter, while the
neighbouring { \it uvw2} and  { \it uvw1} are less sensitive to it. 
The observer frame frequency of the
Ly--$\alpha$ line is shown by the dotted vertical line.
The effective areas of the UVOT filters are also plotted for reference, with
thick dotted lines; the reference scale is reported on the right vertical axis.  
}
\label{lyman}
\end{figure}
%--------------------------------------------------

\section{Optical and X--ray vs $\gamma$--ray flux}

Fig. \ref{uvxg} shows the $\gamma$--ray flux as a function of the flux in the
UVOT {\it uvw2} and {\it b} filters and as a function of the XRT X--ray flux.
From the {\it uvw2} flux we have subtracted the contribution of the accretion disk
(0.55 mJy) and the tail of the Ly--$\alpha$ line (contributing for 0.094 mJy) 
intercepted by this filter (see Fig. \ref{lyman}).
For the $b$ fluxes, we have subtracted a disk contribution of 0.64 mJy.

The total time interval is the same shown in Fig. \ref{lc}, i.e. about 1.5 months.
The $\gamma$--ray flux correlates with both the UV--optical and X--ray fluxes,
and the correlation is more than linear: for a total variation of a factor $\sim$20
of $F_\gamma$, the optical--UV and the X--ray variability amplitude is a factor 
$\sim$3--4.
The vertical grey stripe indicates the flux contribution in the {\it uvw2}
UVOT filter due to the accretion disk emission according to our model plus
the contribution from the tail of the Ly--$\alpha$ line intercepted by the {\it uvw2} filter.

Formally, a least square fit assuming a relation 
$\log F_\gamma = m \log F_{\rm uvw2}+q$
yields $m=1.567$ and $q=-5.026$ (these are the values for the bisector of the correlations
$y=m_1 x+q_1$ and $x=m_2y+q_2$), with a chance probability $P=6.4 \times 10^{-7}$.

The correlation with the X--ray flux, as illustrated in Fig. 2, seems to suggest the presence
of an X--ray constant component contributing at a level of $\sim 5\times 10^{-11}$ 
erg  cm$^{-2}$ s$^{-1}$.
On the other hand, the plotted data refer only to the Nov. -- Dec. 2009 campaign, while
the ``low" state (shown in Fig. \ref{sed}) shows that there are states with lower X-ray and
$\gamma$--ray fluxes (see Tab. \ref{lat} and Tab \ref{xrt}).
The suspected candidate for this behaviour is the SSC component during the Nov.--Dec. 2009
period, as we will discuss in \S 5.1.
Applying a least square fit to the data of Fig. \ref{uvxg} we find
$\log F_\gamma =2.267 \log F_X +17.512$ (again these are the bisector values)
and a chance probability of $P=5\times 10^{-7}$.

The correlated variability in the three bands strongly suggests that the corresponding 
fluxes originate in the same region and by the same population of electrons.
This supports the ``one--zone" models.

The fact that $F_\gamma$ varies more than linearly 
with the --optical--UV and the X--ray
fluxes is interesting and, at first sight, surprising.
Indeed the optical--UV is likely synchrotron emission, while the ~X--ray flux,
having a shape much harder than the optical spectrum, belongs to the same high
energy hump as the $\gamma$--ray emission. 
We then expect that $F_\gamma$ and $F_X$ vary together, with $F_\gamma\propto F_X$.
Furthermore, if the $\gamma$--ray emission is produced by inverse Compton
scattering with broad line photons, produced externally
to the jet -- i.e. it is External Compton (EC) radiation -- it is proportional
to the number of the emitting electrons, as the synchrotron flux. 
Variations of the electron number should then result in $F_\gamma\propto F_{\rm opt}$.
Finally, if the synchrotron Self--Compton model (SSC) process contributes
substantially in the X--ray band, $F_X$ is proportional to the square of the 
number density of the emitting electrons, and then we expect 
$F_X\propto F_\gamma^2$, just the opposite of what observed.
A very similar behaviour has been observed for PKS 1502+106 (Abdo et al. 2010a).

More than linear and even more than quadratic variations 
of the high energy $\gamma$--ray flux with respect to the X--ray flux 
have been already observed in TeV BL Lacs
for relatively short (i.e. weeks) periods of time, as in Mkn 421
(Fossati et al. 2008) and PKS 2155--304 (Aharonian et al. 2009),
but in these cases the X--ray emission belongs to the synchrotron part
of the spectrum, while the $\gamma$--ray flux is probably SSC emission.
Even so, it is not easy to explain a quadratic
relation during the decaying phases of the light curves, because this
implies short cooling timescales (Katarzynski et al. 2005), leading 
Katarzynski \& Walczewska (2010) to propose that the X--ray flux of the varying 
component must be diluted by the flux produced by another region.

In conclusion, we have that the $\gamma$--ray flux varies
more than the flux at lower energies both in 3C 454.3 (and PKS 1502+106) and in the
less powerful BL Lac objects, despite the large differences
in the jet power and in the jet environment leading to different
emission processes for the $\gamma$--ray photons (i.e. SSC for BL Lacs
and EC for 3C 454.3 and other powerful FSRQs).
While these similarities deserve further studies, we note here that 
for powerful blazars we have very short cooling times 
of the emitting electrons, even at low energies: this at least
helps to explain the decaying phases of the light curves,
that are instead a problem for TeV BL Lacs.
For the specific case of 3C 454.3, we will suggest in the following
the likely cause of $F_\gamma$ varying more than $F_X$ and $F_{\rm opt}$,
in terms of an inverse correlation between the dissipated power and the 
magnetic field.

\subsection{Variability and cooling timescales}

Tavecchio et al. (2010) analysed in detail the entire light curve of
3C 454.3 in $\gamma$--rays, including the period of exceptional activity
studied here.
It is found that the minimum doubling timescale 
%for a factor 2 variations 
is between 3 and 
6 hours (in our observer frame). 
This result has been also confirmed by
  Foschini et al. (2010) who studied another flare occurred in early
  April 2010, comparable in flux but observed by Fermi-LAT in pointed mode,
  therefore with denser time coverage and enhanced statistics.
 From Fig. \ref{uvxg} we can see that 
also in X--rays (note the decrease at Dec. 6
and at Dec. 12) the variability timescale is $\sim$1 day or shorter.
This behaviour has two important consequences:
\begin{itemize}
\item 
The size of the emitting region must be compact, $R<ct_{\rm var}\delta/(1+z)
\approx 7\times 10^{15} t_{\rm var,6h}(\delta/20)$ cm.
As remarked in Tavecchio et al. (2010), this challenges models in which the dissipation
region is at tens of parsecs from the black hole 
(see e.g. Jorstad et al. 2010), and instead favours models in which the dissipation occurs much closer.
\item
The cooling time $t_{\rm cool}\le t_{\rm var}$ for electrons emitting at $\gamma$--ray
and X--ray energies.  
This favours models in which the dissipation occurs in a region
where the high density of photons can ensure a rapid cooling.
Again, dissipation close to the black hole is favoured, where the broad line region 
(BLR) can provide a radiation energy density large enough to have 
$t_{\rm cool}<t_{\rm var}$ even for electrons with random Lorentz factors of a few.
\end{itemize}

\subsection{Contribution of the Ly--$\alpha$ and of the accretion disk}
\label{Lyman}

At a redshift of $z=0.859$, the Ly--$\alpha$ emission line from the 
BLR is observed at 2260 \AA, within the band of the {\it uvm2} filter of UVOT
(effective wavelenght $\lambda_{eff}= 2231$ \AA, Poole et al. 2008).

In this source the line is bright (rest frame equivalent width EW = 74 ~\AA\ according
to the observations by Wills et al. 1995), and we wanted to check if it could explain the small
excess in the {\it uvm2} and {\it uvw2} filters that is visible in the SED (Fig. \ref{sed}). 
In Fig. \ref{lyman} we show the 1900--2750 \AA \,portion of the UV spectrum of 
3C 454.3 observed by the \textit{GALEX} 
satellite in October 2008, together with the SED points observed in the UVOT filters and the
corresponding transmission curves.  
% of the uvm2 filter.
All data have been de--reddened with $A_B=0.46$ and $R_{V}= 3.1$,  following Raiteri et al. (2008) 
and assuming the extinction law from Cardelli et al. (1989).
A small mismatch is apparent, but many factors can account for
it, such as systematics in the flux measurement between the two instruments
and  the loose simultaneity of the observations performed with the two telescopes.  
Moreover, a quantitative study on this issue is far beyond the scope of this paper.
Fig. \ref{lyman} shows that the small bump in the portion of the SED observed by
the UVOT filters can be partially due to the  Ly--$\alpha$ emission line
shining the {\it uvm2}, and, at a lower level, the neighbouring filters,
especially {\it uvw2}. This excess was well visible in all our UVOT pointings 
despite the larger level of the non--thermal contribution.
Further, we tried to build a simple model for this portion of the spectrum, as 
the sum of three components:
\begin{itemize}
\item
the Ly--$\alpha$ emission line, modelled as a gaussian profile, with
$\sigma$ and normalisation obtained from a fit  to the \textit{GALEX} spectrum;
\item
the blue bump due to a ``standard" accretion disk, assumed constant
(see \S \ref{modelling});
\item
a variable power law contribution, representing the synchrotron emission from the jet.
\end{itemize} 
With these three components we managed to reproduce the SED profile for each
UVOT observation we considered. 
The excess in the {\it uvm2} filter that we attributed to the Ly--$\alpha$ line 
was compatible, within the uncertainties, with a constant contribution.
This supports Raiteri et al. (2007) when they 
explain the smaller variability range in the UV band w.r.t. the optical
and infrared (IR) as due to the presence of a blue, stable contribution, 
becoming more important (relative to the steep synchrotron non--thermal continuum)
at higher frequencies. 
What we emphasise is the presence of {\it two} constant components: the accretion disk 
and the Ly--$\alpha$ line.

\subsection{The SED of 3C 454.3 at different epochs}

\begin{figure*}
\hskip -0.3cm
\vskip -0.5cm
\psfig{file=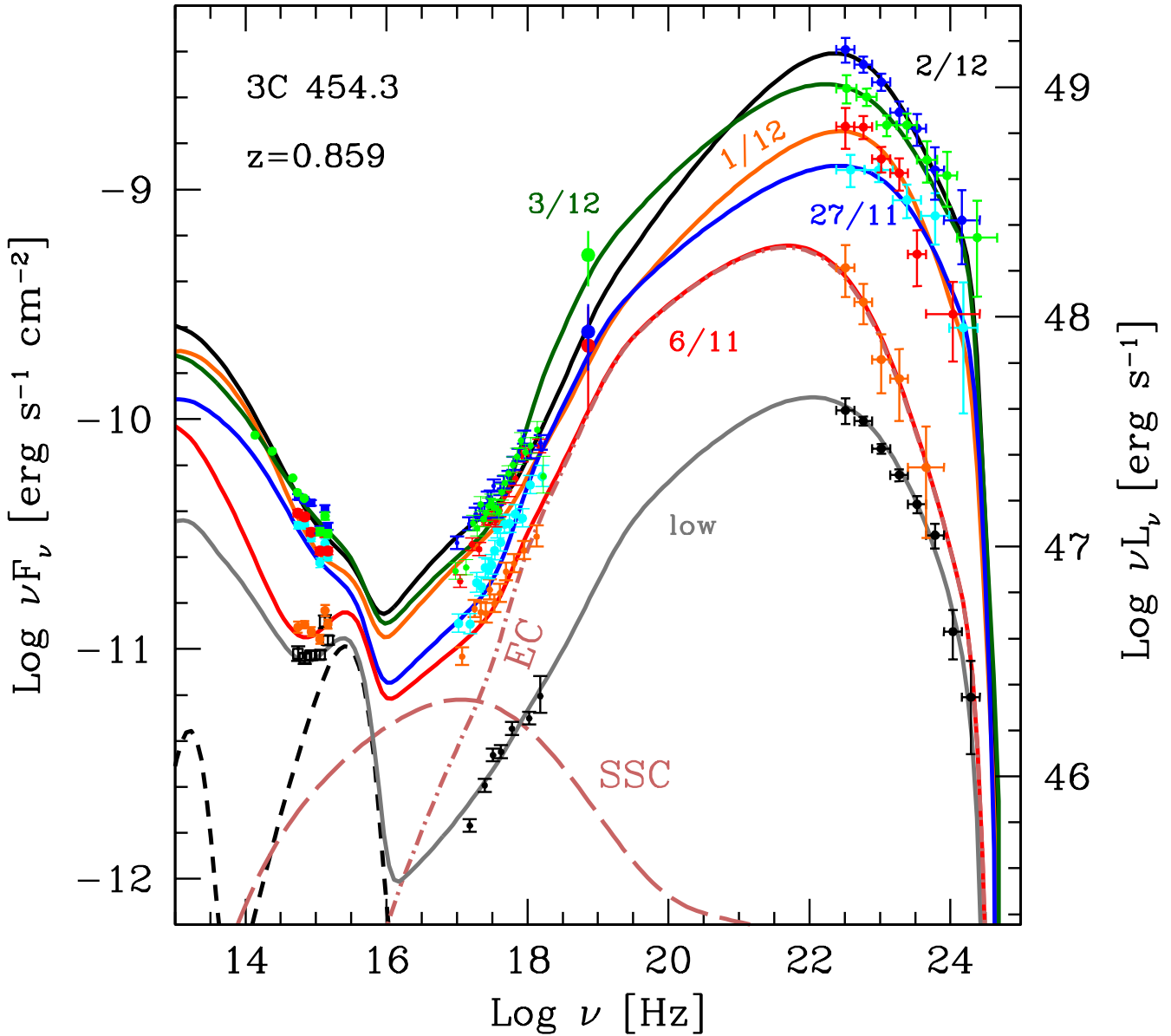,height=18.5cm,width=18.5cm}
\vskip -1.5cm
\caption{The SED of 3C 454.3 at 6 selected epochs: 6 and 27 Nov. 2009,
1, 2, and 3 Dec. 2009 plus the previous ``\emph{low}'' $\gamma$--ray state
(see text).
The different SED are labelled.
We also show the result of our modelling, including the accretion disk
component, its X--ray corona contribution, and the infrared (IR) emission from the torus 
(dashed black lines). 
For the SED of Nov. 6 we show also, for illustration, the contribution of the EC and
SSC components. 
Note that both are necessary to explain the X--ray spectrum, according to our modelling.
}
\label{sed}
\end{figure*}

In Fig. \ref{sed} we show the overall optical to $\gamma$--ray 
SED of 3C 454.3 at 5 different epochs during November and December 2009 and
an additional one (``\emph{low}'') representative of a low $\gamma$--ray state
(see \S \ref{analysis}).
We can see the large variability amplitude of the $\gamma$--ray flux,
whose spectrum is instead remarkably less variable, as noted also in Abdo et
al. (2010b). %Abdo Spettri
Although it is somewhat harder in the ``\emph{low}'' state and on 27 Nov. 2009,
it is always characterised by a slope $\alpha_\gamma>1$ 
[$F(\nu)\propto \nu^{-\alpha}$],
so that the peak of the $\gamma$--ray spectrum is at energies smaller than 100 MeV.
On the other hand the X--ray 0.2--10 keV slope and the {\it Swift}/BAT data points
suggest that the peak cannot be much below 100 MeV.

As discussed above, we can see that the optical--UV flux varies, but less than
the $\gamma$--ray flux. At low states, the accretion disk becomes ``visible"
by flattening the spectrum. 
The presence of the accretion disk emission has been noted by Raiteri et al. (2007)
discussing observations in 2006--2007, that indicate that the disk produced a flux comparable 
with what reported in Fig. \ref{sed}.

As remarked before, also the X--ray flux varied much less than the
$\gamma$--ray one during Nov.--Dec. 2009.
On the other hand, in the ``\emph{low}'' state, the X--ray flux is much lower (also with
respect to the optical), and still very hard.

\subsection{The mass of the black hole of 3C 454.3}

Gu, Cao \& Jiang (2001) estimated $M=4\times 10^9 M_\odot$
for the mass of the black hole of 3C 454.3.
They used the size of the BLR $R_{\rm BLR}$ and the FWHM of 
the H$\beta$ line, equal to 2800 km s$^{-1}$ (Netzer et al. 1995).
To derive the velocity of the clouds, they
multiplied the FWHM value by a factor $f=1.5$
($V=f V_{\rm FWHM}$),  according to McLure \& Dunlop (2001).
The McLure \& Dunlop $f=1.5$ factor was empirically estimated assuming 
that the clouds belong to two components: one moving randomly and
isotropically (that would give $f=\sqrt{3}/2$) 
and one distributed in a disk like geometry.
To derive the black hole mass through $M=R_{\rm BLR} V^2 G^{-1}$,
$R_{\rm BLR}$ was found using the relation of Kaspi et al. (2000):
\begin{equation}
R_{\rm BLR} \sim  4.2 \times 10^{17} 
\left[ {(\lambda L_\lambda)_{5100 \rm \AA} \over 10^{45}
\, \rm erg\, s^{-1} } \right]^{0.700\pm 0.033}\,\, {\rm cm}
\end{equation}
where $(\lambda L_\lambda)_{5100 \rm \AA}$ is the luminosity of the 
continuum at 5100 \AA\ (rest frame, corresponding to 
an observed wavelength of 9480 \AA). 
For high luminosity quasars, the more recent results of Kaspi et al. (2007),
obtained using the CIV $\lambda$1549 line, yield:
\begin{equation}
R_{\rm BLR} \sim 7.8 \times 10^{16} 
\left[ {(\lambda L_{\lambda})_{1350 \rm \AA} \over 10^{45}
\, \rm erg\, s^{-1} } \right]^{0.55\pm 0.04}\,\, {\rm cm}
\end{equation}
The observed luminosity at 2510 \AA\ (corresponding to 1350 \AA, rest frame)
in a period of low state, not much contaminated by the non--thermal continuum,
was $\lambda L_{\lambda}\sim 4\times 10^{46}$ erg s$^{-1}$ 
(from the spectrum taken by the {\it GALEX} satellite reported in Fig. \ref{lyman}).
This yields $R_{\rm BLR} \sim 6\times 10^{17}$ cm.
The FWHM of the CIV line, according to Wills et al. (1995) is 3145 km s$^{-1}$.
Assuming an isotropic velocity field, $f=\sqrt{3}/2$, the mass calculated through
$M=R_{\rm BLR} V^2 G^{-1}$ is $M=3.4\times 10^8 M_\odot$.
Assuming $f=1.5$, instead, we find  $M=1.2\times 10^9 M_\odot$,
in any case smaller than estimated by Gu et al. (2001).
Since the continuum could in fact be contaminated by the non--thermal
beamed contribution, these values should be considered as upper bounds.
Furthermore, given the uncertainties in these relations, this estimate
can give only a rough indication of the black hole mass.
In the following, when modelling the SED, we will use 
$M=5\times 10^8 M_\odot$.
In the general framework of our model, this is the black hole mass 
ensuring an observed minimum variability timescale of $\sim$6 hours,
if the jet opening angle in its conical part is 0.1 rad, as assumed
for all other blazars analysed in Ghisellini et al. (2010).

%--------------------------------------------------
%--------------------------------------------------

\section{Modelling the SED}
\label{modelling}

To model the SED we have used a relatively simple leptonic,
one--zone synchrotron and inverse Compton model.
This model, fully discussed in Ghisellini \& Tavecchio (2009),
has the following main characteristics.

We assume that in a spherical region of radius $R$, located at a distance
$R_{\rm diss}$ from the central black hole, relativistic electrons are injected at
a rate $Q(\gamma)$ [cm$^{-3}$ s$^{-1}$] for a finite time equal to the 
light crossing time $R/c$. 
For the shape of $Q(\gamma)$ we adopt a smoothly broken power law,
with a break at $\gamma_{\rm b}$:
\begin{equation}
Q(\gamma)  \, = \, Q_0\, { (\gamma/\gamma_{\rm b})^{-s_1} \over 1+
(\gamma/\gamma_{\rm b})^{-s_1+s_2} }
\label{qgamma}
\end{equation}

The emitting region is moving with a  velocity $\beta c$
corresponding to a bulk Lorentz factor $\Gamma$.
We observe the source at the viewing angle $\theta_{\rm v}$ and the Doppler
factor is $\delta = 1/[\Gamma(1-\beta\cos\theta_{\rm v})]$.
The magnetic field $B$ is tangled and uniform throughout the emitting region.
We take into account several sources of radiation externally to the jet:
i) the broad line photons, assumed to re--emit 10\% of the accretion luminosity
from a shell--like distribution of clouds located at a distance 
$R_{\rm BLR}=10^{17}L_{\rm d, 45}^{1/2}$ cm;
ii) the IR emission from a dusty torus, located at a distance
$R_{\rm IR}=2.5\times 10^{18}L_{\rm d, 45}^{1/2}$ cm;
iii) the direct emission from the accretion disk, including its X--ray corona.
Furthermore we take into account 
the starlight contribution from the inner region of the host galaxy
and the cosmic background radiation, but these photon sources
are unimportant in our case.
We instead neglect, for simplicity, the contribution of a possible
population of intra--cloud free electrons. This contribution
would be important if the optical depth $\tau$ of these electrons
became greater than the covering factor of the emitting line clouds, i.e. $\tau>0.1$,
a rather large value. These free electrons would scatter and re--isotropize the entire
disk flux, providing IR radiation, besides optical and UV.
The final effect would be to broaden the high energy hump, especially below its
peak.

All these contributions are evaluated in the blob comoving frame, where we 
calculate the corresponding inverse Compton radiation from all these contributions, 
and then transform into the observer frame.
For simplicity, the numerical code assumes an abrupt cut-off for the scattering cross section, 
equal to the 
Thomson one for  $\gamma h\nu_{\rm Ly-\alpha}/m_{\rm e} c^2 < 1$ and zero above. This approximation 
provides acceptable results in the calculation of the IC scattering of the BLR 
(e.g. Tavecchio \& Ghisellini 2008) and starts to fail only at the largest energies probed 
by LAT, at which scatterings in the full Klein-Nishina regime start to be  important 
(see also Georganopoulos Kirk \& Mastichiadis 2001; Moderski et al. 2005 for a discussion on 
the importance of the full Klein-Nishina cross section).

We calculate the energy distribution $N(\gamma)$ [cm$^{-3}$]
of the emitting particles at the particular time $R/c$, 
when the injection process ends. 
Our numerical code solves the continuity equation which includes injection, 
radiative cooling and $e^\pm$ pair production and reprocessing. 
Our is not a time dependent code: we give a ``snapshot" of the 
predicted SED at the time $R/c$, when the particle distribution $N(\gamma)$ 
and consequently the produced flux are at their maximum.

For 3C 454.3 the radiative cooling time of the 
particles is short, shorter than $R/c$ even for low energetic particles.
This implies that, at lower energies, the $N(\gamma)$ distribution is proportional
to $\gamma^{-2}$, while, above $\gamma_{\rm b}$, $N(\gamma)\propto \gamma^{-(s_2+1)}$.
The electrons emitting most of the observed radiation have energies 
$\gamma_{\rm peak}$ which is close to $\gamma_{\rm b}$ 
(but these two energies are not exactly equal, due to the curved 
injected spectrum).

The accretion disk component is calculated assuming a standard
optically thick geometrically thin Shakura \& Sunjaev (1973) disk.
The emission is locally a black body, with a temperature profile
given e.g. in Frank, King \& Raine (2002).
Since the optical--UV is the sum of the accretion disk and the jet non--thermal
components, there is some degeneracy when deriving the black hole mass and
the accretion rate, even in the lowest state, for which the disk is
best visible.
On the other hand we can take as a guideline the value of the black hole mass
derived in \S 4.1, implying then a very large accretion rate.
We also note that in modeling the optical-UV region one should also consider 
other thermal components, such as the other emission lines and the Balmer continuum from the BLR. 
However, all these components are expected to be much less bright than the Ly$-\alpha $ emission line, 
included in the fit. It is also possible that the disk emission spectrum is
different than that expected by the Shakura \& Sunjaev model: however we use the latter, built on robust theoretical basis.

We model at the same time the thermal disk (and IR torus) radiation 
and the non--thermal jet--emission.
The link between these two components is given by the amount of 
radiation energy density (as seen in the comoving frame of the emitting
blob) coming directly from the accretion disk or reprocessed by the BLR and
the IR torus.
This radiation energy density depends mainly on $R_{\rm diss}$, but
not on the adopted accretion rate or black hole mass
(they are in any case chosen to reproduce the observed thermal disk 
luminosity).

By estimating the physical parameters of the source we can calculate
the power that the jet carries  in the form of radiation
($P_{\rm r}$), magnetic field ($P_{\rm B}$), relativistic electrons
($P_{\rm e}$) and cold protons ($P_{\rm p}$) assuming one proton per
electron.
These powers are calculated according to:
\begin{equation}
P_{\rm i} \, =\, \pi R^2 \Gamma^2 c U^\prime_{\rm i}
\label{power}
\end{equation}
where $U^\prime$ is the energy density of the $i_{\rm th}$ component
in the comoving frame.

\subsection{Constraints on the model parameters}
\label{constraints}

We briefly discuss the constraints that we can put on our modelling.

\begin{itemize}
\item 
The size $R$ of the emitting region must be compact,
in order to vary with the observed short timescale.
In the adopted model, this corresponds to a location of a jet dissipation
region $R_{\rm diss}\sim ct_{\rm var}\delta/[(1+z)\psi]\sim 
7\times 10^{16} t_{\rm var,6h}(\delta/20)/\psi_{-1}$ cm,
if the jet opening angle is $\psi=0.1 \psi_{-1}$.

\item The short variability timescale and the large
disk luminosity imply $R_{\rm diss}<R_{\rm BLR}$: the dissipation
occurs within the BLR.

\item If so, the relevant radiation energy density is provided by the BLR photons,
with energy density 
$U^\prime_{\rm BLR}\sim f_{\rm BLR} L_{\rm d} \Gamma^2 /(4\pi R_{\rm BLR}^2 c)$.
Setting $f_{\rm BLR}=0.1$ and using 
$R_{\rm BLR}=10^{17}L_{\rm d, 45}^{1/2}$ cm, the resulting 
$U^\prime_{\rm BLR}\sim \Gamma^2/(12\pi)$, i.e. the dependence on $L_{\rm d}$
and $R_{\rm BLR}$ drops out.
As a consequence the Lorentz factor of the electrons cooling
in a time $R/c$ is ($U_B$ and $U^\prime_{\rm syn}$ are the magnetic and the 
synchrotron energy densities):
\begin{eqnarray}
\gamma_{\rm cool} &=& { 3 m_{\rm e}c^2 \over 
4\sigma_{\rm T} U^\prime_{\rm BLR} R [1+(U_B+U^\prime_{\rm syn})/U^\prime_{\rm BLR}] }
\nonumber \\
&\sim& {9 \over (\Gamma/20)^2 R_{16}[1+(U_B+U^\prime_{\rm syn})/U^\prime_{\rm BLR}] } 
\end{eqnarray}
Therefore the radiative cooling is efficient for all but the lowest
energy electrons.

\item
The bulk Lorentz factor $\Gamma$ and the Doppler factor $\delta$
are constrained by the observed superluminal motion at the VLBI scale.
Jorstad et al. (2005) find different components moving
with different apparent velocities $\beta_{\rm app}$, 
from a few to more than 20, for observations
performed between 1998 and 2001, resulting in different bulk Lorentz factors (from 10 to 25) 
and different orientations (viewing angles from 0.2 to 4 degrees), resulting in different
Doppler factors ($\delta$ from 14 to 30).
More recent observations by Lister et al. (2009) found
$\beta_{\rm app}=14.2\pm 0.8$, consistent with the previous findings.

\item
The soft slope of the $\gamma$--ray spectrum and the hard slope of
the X--ray spectrum constrain the peak of the high energy component $h\nu_{\rm c}$
of the SED to lie below 100 MeV (but close to this value).
For the EC process, the peak is made by the most relevant electrons
scattering the Ly--$\alpha$ seed photons.
They must have a Lorentz factor $\gamma_{\rm peak}$ given by
\begin{eqnarray}
\nu_{\rm c} &\sim& \gamma_{\rm peak}^2 \nu_{Ly\alpha} {\Gamma\delta\over 1+z} 
\nonumber \\
&\to&
\gamma_{\rm peak} \sim 210 \, \left( {h\nu_{\rm c}\over 100\, {\rm MeV}}\right)^{1/2}
\left({\delta\Gamma\over 400} \right)^{1/2}
\end{eqnarray}
This implies that the emission at and above 
the $\gamma$--ray peak is in fast cooling
(i.e. that the cooling time is shorter than $R/c$).

\item
Since the bolometric luminosity is dominated by the $\gamma$--ray emission, 
this limits the injected power in the form of relativistic electrons.
The fast cooling regime ensures that almost all the injected power is 
converted into radiation.
We then have $P^\prime_{\rm inj}\approx L_\gamma/\delta^4$.

\item
In general, the value of the magnetic field in the emitting region is derived
by the level of the synchrotron emission.
However the peak of the synchrotron spectrum, and hence its bolometric luminosity,
lies in the unobserved mm--far IR frequency range. 
A more robust constrain comes from the soft X--ray band, where 
we have the contribution of the EC and SSC components.
Their spectrum is different (the SSC is softer).
Increasing the magnetic field $B$ increases the SSC component.
The $B$--field is then found by reproducing the observed flux and shape of 
the X--ray spectrum.

\item
The observed soft shape of the $\gamma$--ray and optical--UV spectra
(once accounting for the contribution of the accretion emission, 
important in the low states) constraints the index $s_2$ of the energy
distribution of the injected electrons.
Since the cooling is almost complete (i.e. almost all electrons cool in $R/c$),
the shape of the emitting distribution, below $\gamma_{\rm peak}$, 
is $\propto \gamma^2$, almost independent of $s_1$ (if $s_1<2$).
On the other hand, the details of the curvature around $\gamma_{\rm peak}$
and the distribution below $\gamma_{\rm cool}$ do depend on $s_1$.
This index is then found by reproducing the X--ray spectrum (and the 
{\it Swift}/BAT point).

\end{itemize}

% =============================================
\begin{table}
\begin{center}
\begin{tabular}{lllllll}
\hline
\hline
                     &\emph{low}     &06/11 &27/11  &01/12  &02/12  &03/12   \\
\hline
$R_{\rm diss}$       &132     &156   &174    &150    &173    &174      \\
$P^\prime_{\rm inj}$ &0.012   &0.054 &0.083  &0.09   &0.145  &0.13   \\
$B$                  &6.35    &5.79  &4.57   &5.24   &4.43   &4.49     \\
$\Gamma$             &15      &15    &17     &18     &19.6   &19.7          \\
$\gamma_{\rm b}$     &400     &340   &800    &600    &350    &430        \\
$\gamma_{\rm max}$   &2.0e3   &3.4e3 &2.8e3  &3.0e3  &2.8e3  &3.0e3     \\
$s_1$                &1.2     &1.4   &1.45   &1.2    &0.8    &1.3         \\
$s_2$                &3.4     &4.0   &3.9    &4.3    &3.5    &3.4         \\
\hline
$R_{\rm diss}/R_{\rm s}$ &880 &1040  &1160   &1000   &1150   &1160   \\
$R_{\rm blob}$       &13.2    &15.6  &17.4   &15.0   &17.6   &17.7     \\
$\gamma_{\rm peak}$  &149     &95.8  &206.6  &203    &176    &144            \\
$\delta$             &24.5    &24.5  &26.4   &27.3   &28.4   &28.5                 \\
$t_{\rm var}$        &5.0     &5.9   &6.1    &5.1    &5.7    &5.8             \\
$\log P_{\rm r}$     &45.35   &45.98 &46.33  &46.42  &46.72  &46.66   \\
$\log P_{\rm p}$     &47.08   &47.97 &48.11  &48.01  &48.09  &48.37   \\
$\log P_{\rm e}$     &44.62   &45.40 &45.51  &45.51  &45.63  &45.76   \\
$\log P_{B}$         &45.84   &45.84 &45.84  &45.88  &45.94  &45.96   \\
$\dot M_{\rm out}$   &0.14    &1.08  &1.33   &1.01   &1.08   &2.09    \\
\hline
\end{tabular}
\caption{For all models we have assumed a viewing angle $\theta=1.8^\circ$ 
and a bolometric luminosity of the accretion disk 
$L_{\rm disk}=6.75\times 10^{46}$ erg s$^{-1}$.
The black hole mass is assumed to be $M=5\times 10^8M_\odot$, so
the disk luminosity is the 90\% of the Eddington value,
and the mass accretion rate is $\dot M_{\rm in}=14.78 M_\odot$ yr$^{-1}$
(assuming an efficiency $\eta=0.08$).
The size of the emitting blob, for the assumed distances 
$R_{\rm diss}$ from the black hole,
is always $\psi R_{\rm diss}$, with $\psi=0.1$.
We assume that the 10\% of the disk luminosity
is reprocessed by the BLR and re--emitted as broad lines
at a distance $R_{\rm BLR}=8.2\times 10^{17}$ cm for all models.
Another 30\% of the disk luminosity is reprocessed and re--emitted as
IR radiation by a torus, located at $R_{\rm IR} \sim 2\times 10^{19}$ cm.
The minimum observed variability timescale $t_{\rm var}$ is in hours and
is defined as $t_{\rm var}=R(1+z)/(c\delta)$. 
Powers are in units of erg s$^{-1}$.
Size and distances in units of $10^{15}$ cm. 
Magnetic field $B$ in Gauss.
Outflow mass rates $\dot M_{\rm out}$ in $M_\odot$ yr$^{-1}$.
}
\label{para}
\end{center}
\end{table}
% =============================================

\section{Results}

\begin{figure}
\hskip -0.3cm
\vskip -0.3cm
\psfig{file=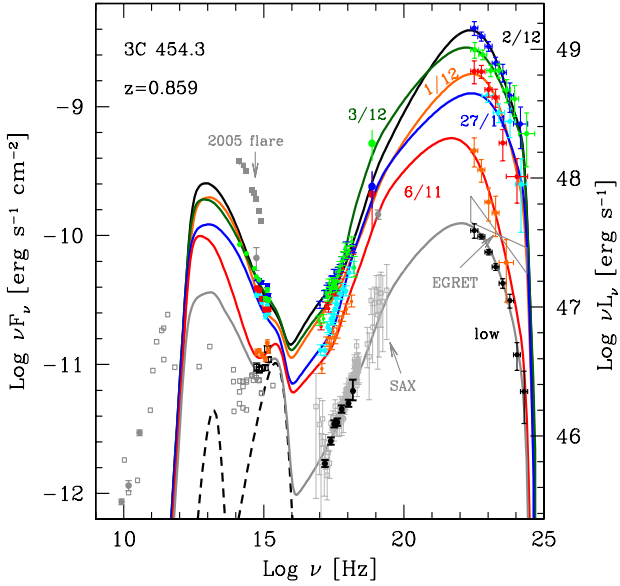,height=9cm,width=8.5cm}
\vskip -0.7cm
\caption{Same as Fig. \ref{sed}, but extending to radio frequencies
and showing also archival data (in light grey), including the optical 
fluxes achieved during the 2005 optical flare, the 5--6 June 2000
{\it Beppo}SAX spectrum (Tavecchio et al. 2002) and the EGRET spectrum of Jan 1992 
(Nandikotkur et al. 2007).
Note that 
i) during the optical flare of 2005 the optical flux reached a brighter
state than in Nov--Dec 2009; 
ii) the low {\it Fermi}/LAT state is at a lower level than detected by EGRET and 
iii) that the X--ray flux and spectrum detected by {\it Swift}/XRT 
for the ``\emph{low}'' state  superpose exactly to the {\it Beppo}SAX data. 
}
\label{sedall}
\end{figure}

Fig. \ref{sed} shows the results of our modelling for the 6 simultaneous SEDs 
for different epochs, while Tab. \ref{para} lists the used parameters.
Fig. \ref{sedall} shows the same SEDs and models, but extending the frequency axis
to the radio band and showing some archival data, to put the SEDs analysed here
in contest. 

The model successfully reproduces the data from the optical to the $\gamma$--ray 
band, using the guidelines explained in \S \ref{constraints}. %the previous sub--section.
The large variations of the $\gamma$--ray luminosity are obtained by mainly 
varying the power injected in relativistic electrons (by a factor 10 from the ``\emph{low}''
to the highest state, and by a factor 3 in the restricted period from Nov 6 to Dec 2),
and the bulk Lorentz factor (from 15 to 20). 
These are the main changes.
Besides them, there are minor changes in the location of the dissipation region
(hence, in the size of the emitting region) by less than a factor 1.4, and in the
parameters of the injected electron distribution (i.e. $\gamma_{\rm b}$ changes 
by a factor $\sim2$, and also $s_1$ changes somewhat). 
These changes are required to fit the X--ray spectrum, while the changes in the
$s_2$ parameter are necessary to account for the (small) changes of the 
$\gamma$--ray and the optical--UV spectra.
Although minor, these changes have a relatively strong impact on the total
jet power, because $\gamma_{\rm b}$ and $s_1$ control the total amount
of electrons present in the source, thus also the number of cold protons,
if we assume one proton per electron.
Indeed, the maximum value of $P_{\rm p}$ is obtained on Dec. 3, not at the
maximum of the $\gamma$--ray flux (occurring at Dec. 2, when there is also 
the maximum value of $P^\prime_{\rm inj}$).

As explained in \S \ref{constraints}, the value of the magnetic
field is mainly derived to adequately fit the X--ray spectrum, rather
than the synchrotron optical--UV spectrum and flux level.
Fig. \ref{sed} shows the contributions of the SSC and EC components for
the SED of Nov 6.
They intersect at $\sim$1 keV, with the SSC dominating below, and the EC above.
A similar decomposition has been adopted in Vercellone et al. (2010)
for another epoch
(see their Fig. 19, showing also the contribution of the accretion disk
radiation to the optical--UV flux). 
% Although these two components are assumed to be produced by electrons 
% belonging to the same distributions, the energies of the electrons producing
% X--rays by the SSC and EC are largely different.
% One may then expect a different variability behaviour below and above 1 keV,
% to be checked through X--ray data extending for a time longer than for
% the data shown in Fig. \ref{sed}.

Fig. \ref{sedall} shows that the during the optical flare of 2005 the optical 
flux reached a brighter state than in Nov--Dec 2009.
There was no $\gamma$--ray facility in orbit in 2005, so we do not know the
$\gamma$--ray flux during the optical flare, interpreted by Katarzynski \& Ghisellini
(2007, see also Ghisellini et al. 2007) as a dissipation event occurring 
relatively close to the black hole, where the magnetic energy density can be of 
the same order of the radiation energy density as seen in the comoving frame. 
In this case $L_{\rm syn}\sim L_{\rm EC}\sim L_\gamma$, and we do not need 
an extraordinary increase of the total power budget to explain this
extraordinary optical flare that can occur without a corresponding large 
increase of the $\gamma$--ray flux (and the bolometric one; in this sense
the model can be thought as ``economic", i.e. assuming the lowest possible
power budget).
On the contrary, the 2009 $\gamma$--ray flare does change the bolometric
luminosity by a large factor, without inducing a correspondingly dramatic
increase of the optical flux.
The results of the modelling presented here explain this behaviour by having
a larger power dissipated in regions more distant than in the 2005 flare,
with relatively smaller magnetic field and larger bulk Lorentz factors.

Fig. \ref{sedall} also shows that the ``\emph{low}'' {\it Fermi}/LAT state is 
lower than the $\gamma$--ray flux detected by EGRET in Jan. 1992 and that 
the X--ray flux and spectrum detected by {\it Swift}/XRT for the
``\emph{low}'' state  superpose exactly to the {\it Beppo}SAX data of 5--6
June 2000 (discussed in Tavecchio et al. 2002).

%--------------------------------------------------
%--------------------------------------------------

%--------------------------------------------------
\begin{figure}
\vskip -0.6 cm
\hskip -0.5 cm
\psfig{file=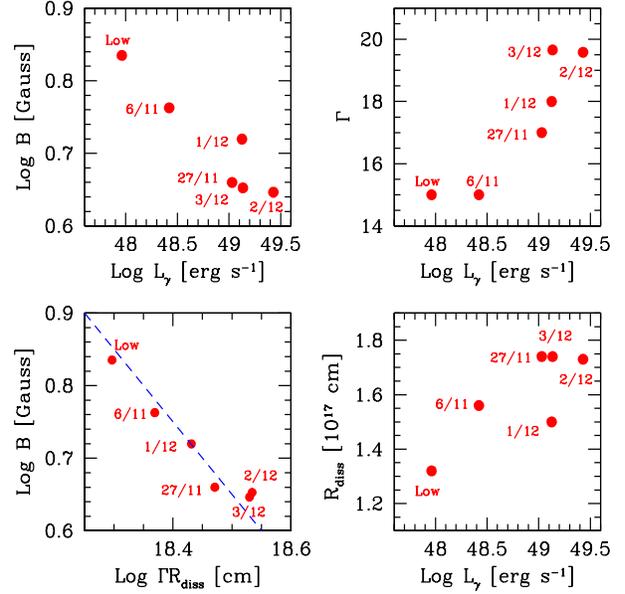,height=9.5cm,width=9cm}
\vskip -0.5cm
\caption{
Top: the magnetic field $B$ (left)
and the bulk Lorentz factor $\Gamma$ (right)
as a function of the $\gamma$--ray luminosity $L_\gamma$ for the 6 different
SED plotted in Fig. \ref{sed}.
Bottom left: the magnetic field $B$ as a function of $\Gamma R_{\rm diss}$.
Bottom right: $R_{\rm diss}$ as a function of $L_\gamma$.
}
\label{figpara}
\end{figure}
%--------------------------------------------------

\subsection{$\gamma$--ray vs X--ray and $\gamma$--ray vs optical flux correlations}

In \S 3 we showed that the $\gamma$--ray flux varies almost quadratically with the
optical and X--ray fluxes.
In the framework of synchrotron + External Compton model, modest variations of
the synchrotron flux accompanied by large variations of the EC flux 
imply that the magnetic field must vary oppositely
with the observed bolometric power (i.e. the $\gamma$--ray luminosity, in this case).
Another way to have more than linear variations (but not quadratic)
is to vary $\Gamma$, that would result in $L_{\rm EC} \propto L_{\rm syn}^{3/2}$
(this relation, valid considering frequency integrated
luminosities, remains valid in restricted energy ranges if the EC and
synchrotron peaks do not move).

Fig. \ref{figpara} illustrates the specific way of our modelling to have 
the observed $L_\gamma \propto L_{\rm opt}^2$ correlation.
The figure shows how $B$, $\Gamma$ and $R_{\rm diss}$ depend on the observed $L_{\gamma}$,
and how $B$ depends on the product $\Gamma R_{\rm diss}$.
We can see that $B$ inversely correlates with $L_\gamma$, while $\Gamma$ and $R_{\rm diss}$,
instead, correlate positively.
These makes $B$ to be almost perfectly proportional to $(\Gamma R_{\rm diss})^{-1}$.
Since $P_{\rm B}\propto [R_{\rm diss} \Gamma B]^2$, we have that the Poynting
flux at $R_{diss}$ is approximately constant for all the considered states.

To summarise: the large variation in the $\gamma$--ray flux accompanied by much more
modest variations of the optical are explained by a magnetic field that decreases
when the observed $L_\gamma$ increases.
This is accomplished, in our model, by having larger $R_{\rm diss}$ and $\Gamma$
for larger $L_\gamma$ and jet powers, and by having instead a quasi--constant
Poynting flux in the emission region. 
By the same argument we can explain the $F_\gamma$--$F_X$ correlation,
since the contribution of the SSC component to the X--rays becomes relatively
more important for {\it lower} states, making the SSC X-ray flux to decrease much
less than the $\gamma$--ray one.

However, the quasi constancy of $P_{\rm B}$ may be partly a coincidence, 
and not the outcome of a fundamental process.
This is because it is very likely that the Poynting flux $P_{\rm B,0}$
at the start of the jet is the prime cause of the acceleration of the jet itself.
In this case, since we find that the total jet power does change, $P_{\rm B, 0}$ 
should change as well, and yet have the same value when arriving 
at $R_{\rm diss}$.
What can happen is that a stronger $P_{\rm B,0}$ implies
a larger final $\Gamma$, achieved at a larger distance $z$
(i.e. we can have $\Gamma\propto z^{1/2}$ as in Vlahakis et al. 2004).
This means that a larger $P_{\rm jet}\sim P_{\rm B,0}$ is achieved, but since 
both $R_{\rm diss}$ and $\Gamma$ are larger, $P_{\rm B}$ at the dissipation region
may vary much less than $P_{\rm B,0}\sim P_{\rm jet}\propto L_\gamma$.

%--------------------------------------------------
\begin{figure}
\vskip -0.6 cm
\psfig{file=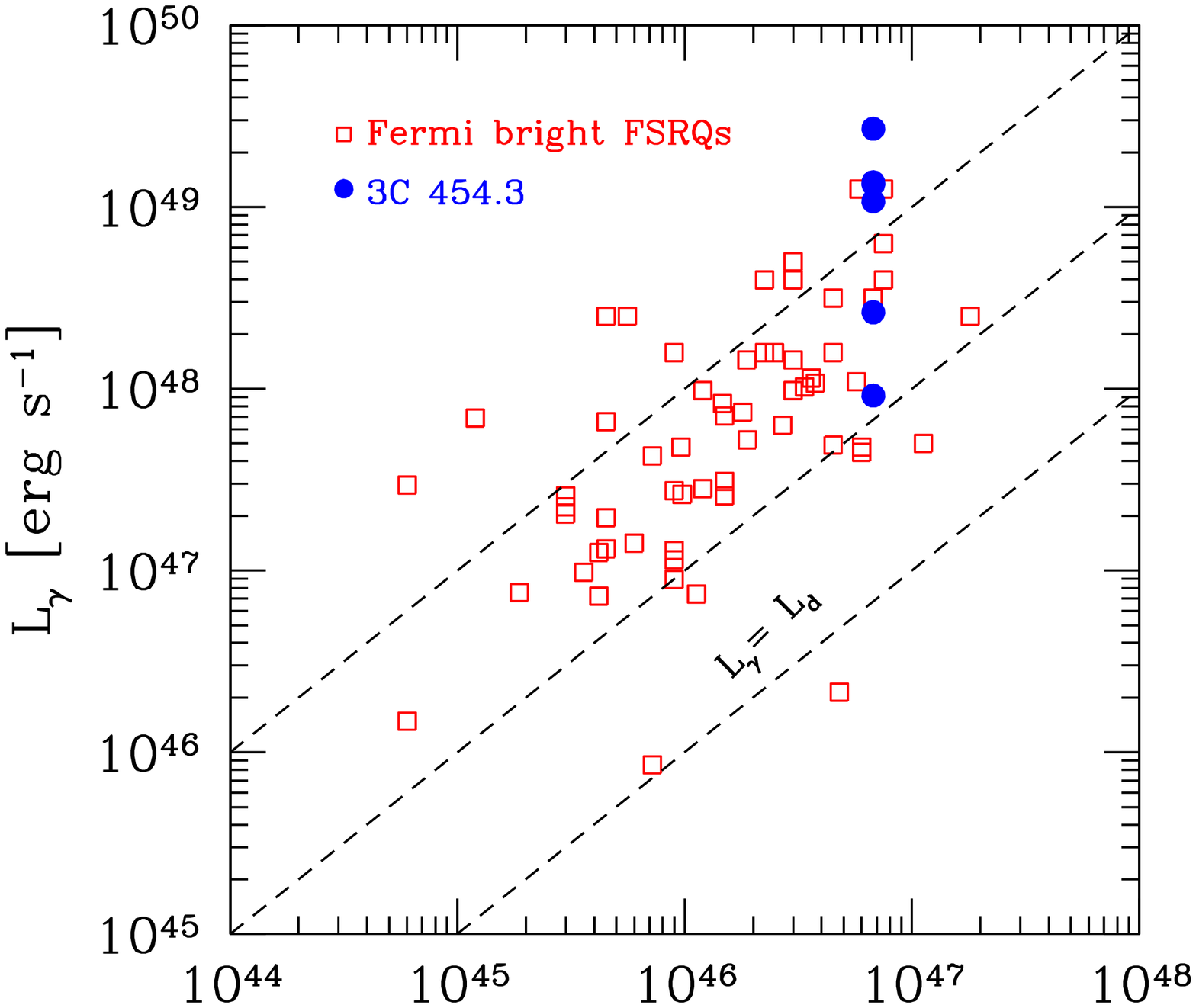,height=8cm,width=9cm}
\vskip -1.3 cm
\psfig{file=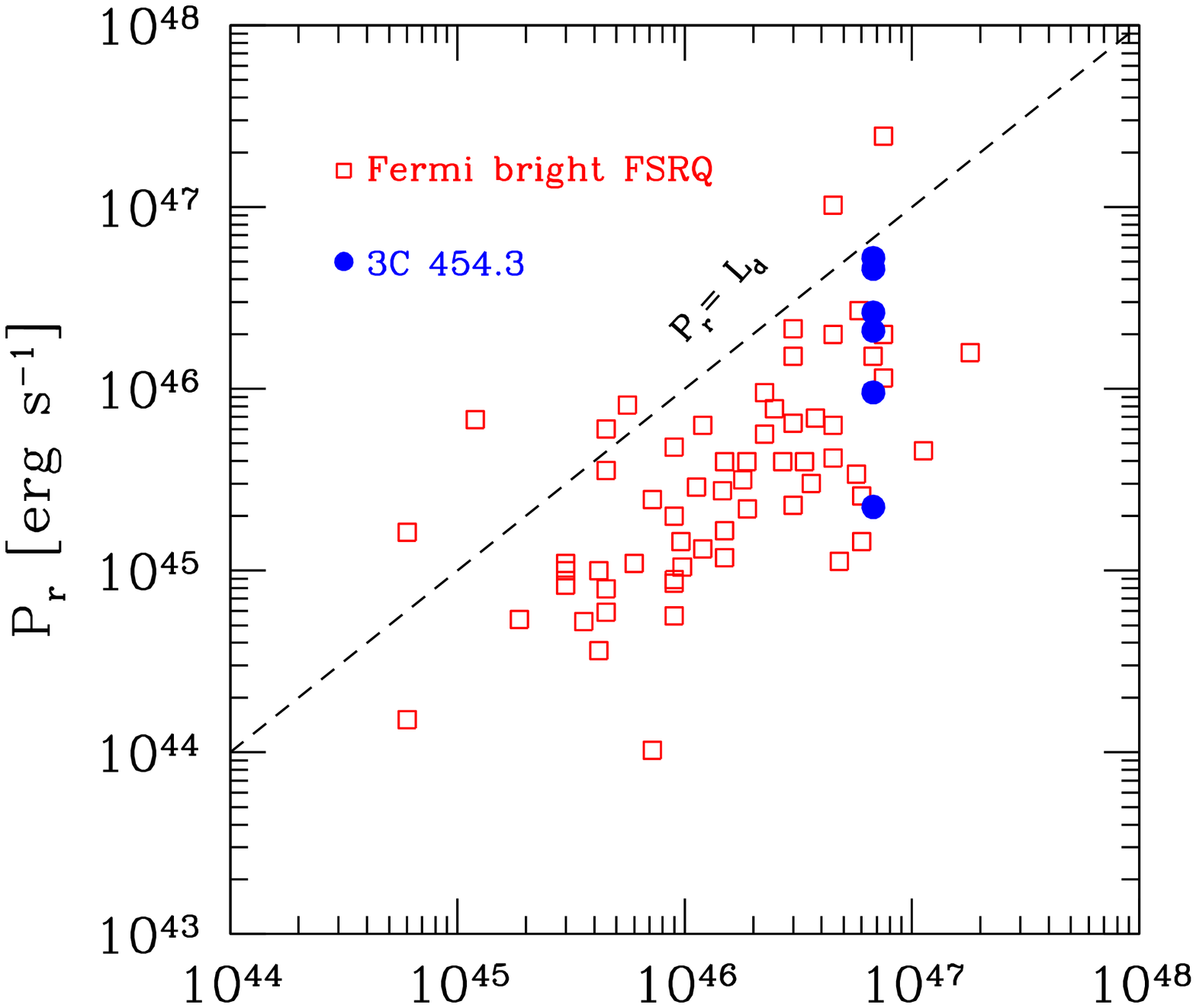,height=8cm,width=9cm}
\vskip -1.3 cm
\psfig{file=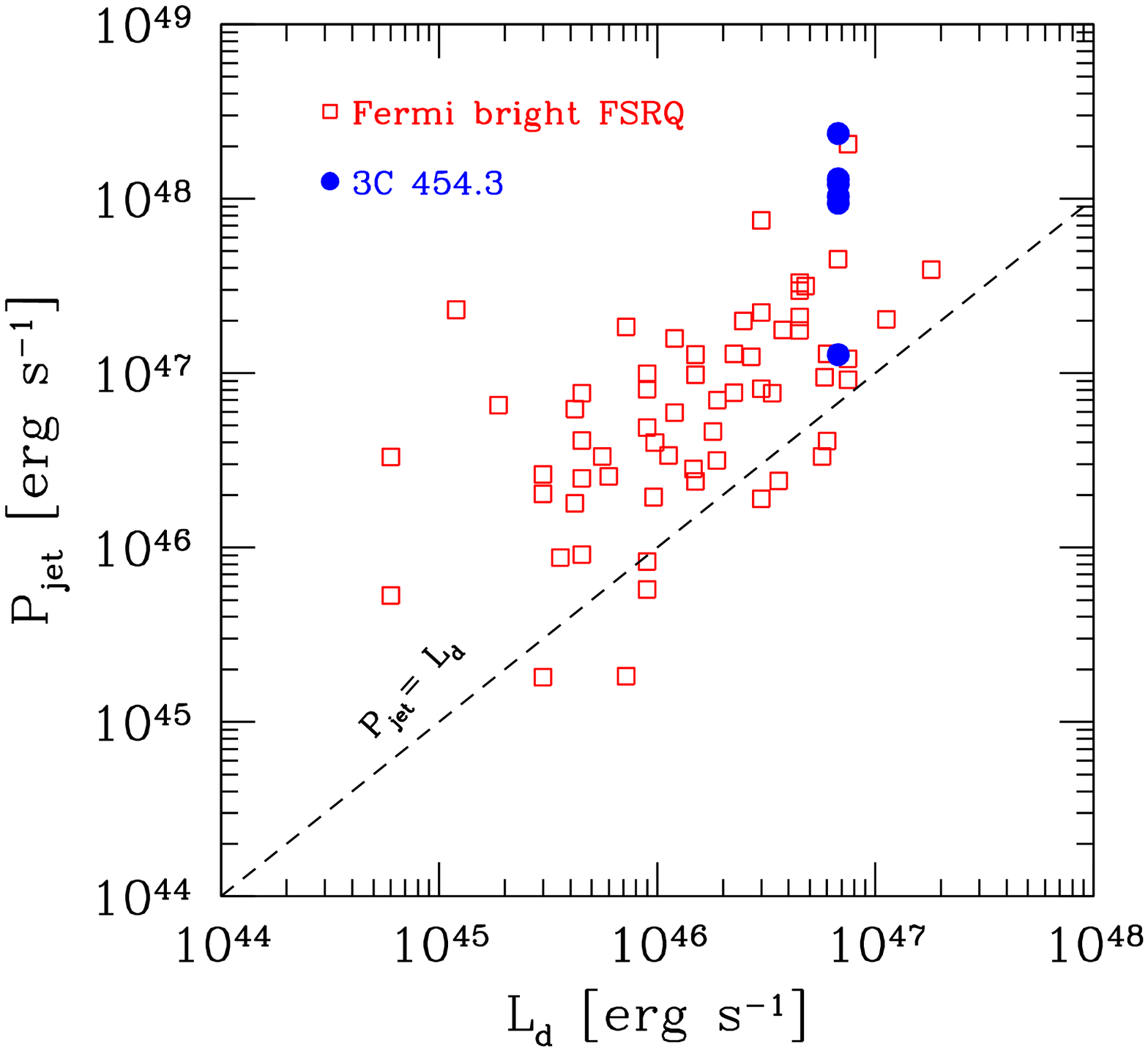,height=8cm,width=9cm}
\vskip -0.5cm
\caption{
The observed $\gamma$--ray luminosity $L_\gamma$ (top panel)
the power $P_{\rm r}$ spent by the jet to produce the radiation we see
(mid panel) and the total jet power $P_{\rm jet}$ (botton panel)
as a function of the accretion disk luminosity $L_{\rm d}$.
The filled circles correspond to the 6 states of 3C 454.3
analysed in this paper, while the empty squares
correspond to all the FSRQs detected in the first 3--months of the {\it Fermi}/LAT
all sky survey analysed by Ghisellini et al. 2010. % General Properties
}
\label{ldlg}
\end{figure}
%--------------------------------------------------

\subsection{The power of the jet}

It is interesting to investigate if the extraordinary $\gamma$--ray 
flare corresponds to a real enhancement of the power carried by the jet
or if originates by an enhanced efficiency in converting the jet
power into radiation, or if is due to an increased Doppler factor $\delta$,
making the observed flux to increase even if the jet power remains constant.
The nearly model--independent quantity connected to the jet power is 
$P_{\rm r}$, the power that the jet spends to produce the radiation we see.
Using $U^\prime_{\rm rad}=L^\prime/(4\pi R^2 c)$, we can re--write
Eq. \ref{power} as
\begin{equation}
P_{\rm r}  \, =\,  L^\prime {\Gamma^2 \over 4} \, =\, L {\Gamma^2 \over 4 \delta^4}
\, \sim \, L {1 \over 4 \delta^2}
\end{equation} 
where $L$ is the total observed non--thermal luminosity
($L^\prime$ is in the comoving frame) and $U^\prime_{\rm rad}$ is the 
radiation energy density produced by the jet (i.e.
excluding the external components).
The last equality assumes $\theta_{\rm v}\sim 1/\Gamma$.
This quantity is almost model--independent,
since it depends only on the adopted $\delta$, that can be estimated 
also by other means, namely superluminal motions.
In Tab. \ref{para} we list the values of $P_{\rm r}$ for the
different states. 
The maximum is on Dec 2, the day of the maximum $L_\gamma$, when
$P_{\rm r}=5.2\times 10^{46}$ erg s$^{-1}$.
This value can be compared with the accretion luminosity, that we assumed
constant during the entire period and equal to 
$L_{\rm d}=6.75\times 10^{46}$ erg s$^{-1}$.
The power $P_{\rm r}$ can be considered as a very robust lower limit to the total
jet power $P_{\rm jet}=P_{\rm B}+P_{\rm e}+P_{\rm p}$.
If the entire $P_{\rm jet}$ is used to produce the radiation we see
(i.e. $P_{\rm jet}=P_{\rm r}$), then the jet would necessarily decelerate and
there would be no VLBI superluminal knots, nor any extended radio emission.

Tab. \ref{para} lists also the value if $P_{\rm B}$ and $P_{\rm e}$.
They are smaller than $P_{\rm r}$.
The fact that $P_{\rm e}<P_{\rm r}$ may seem strange at first sight, since 
the radiation is produced by the electrons, so how can they have less power
that what they transform in radiation?
The answer is that the radiative cooling time is shorter than $R/c$, so 
electrons need to be injected continuously, at least for a time $R/c$.
To estimate $P_{\rm e}$ we count electrons (and their energy) 
forming the  $N(\gamma)$ [cm$^{-3}$] energy distribution after solving
the continuity equation, not the ones injected.
An example can clarify this point: assume to inject the same energy
distribution $Q(\gamma)$ [cm$^{-3}$ s$^{-1}$]. 
In case A the radiative cooling $\dot\gamma$ is strong, in case B is weak.
The particle density $N(\gamma)$ is always proportional to
$[\int Q(\gamma)d(\gamma)] /\dot \gamma$, therefore in case A $N(\gamma)$ is
smaller than in case B.
This despite the fact that in case A we have produced more radiation,
since the cooling is stronger.

This implies that there must be a ``reservoir" of power able to energise
electrons that in turn produce $P_{\rm r}$.
The simplest solution is to assume that this ``reservoir" is
provided by protons.
The listed values of $P_{\rm p}$ assume one proton per electron.
The implication is that electron--positron pairs cannot be
energetically dominant, even if we cannot exclude that they outnumber
primary electrons (but not by a large factor).
For a more detailed discussion about this point see 
Sikora \& Madejski (2000) and Celotti \& Ghisellini (2008).
We summarize here the arguments in Celotti \& Ghisellini (2007).
If pairs are dynamically important, they must outnumber protons
and also the emitting leptons by a large factor.
Although possible in principle, this solution poses the problem of
the origin of this large number of pairs.
If they have been created in the same emitting region (e.g. by 
$\gamma$--$\gamma$ collisions), then we should see i) a cutoff in the spectrum, 
and ii) some reprocessed radiation in X--rays (since the pairs are born relativistic).
The level of this reprocessed radiation must be of the same order of the
absorbed luminosity (i.e. it is large and should be well visible).
If, instead, the pairs are created from the start, we can calculate
how many we need. In powerful sources (as 3C 454.3) this number corresponds
to an initial pair optical depth greater than unity. If initially cold,
the pairs quickly annihilate, and the surviving ones are not enough (to
be dynamically important). If they are hot, they emit. At the start of the
jet, close to the accretion disk, the radiation energy density and 
the magnetic field make them to cool very rapidly. So they become cold,
and annihilate (besides producing radiation we do not see).

Assuming then that $P_{\rm p}$ is a good proxy for the real $P_{\rm jet}$ 
we see that during Nov and Dec 2009 the jet power varied only by a factor
of 2, while $L_\gamma$ varied by a factor $\sim$10.
This implies that the jet became more efficient in transforming
its power into radiation, rather than becoming more powerful.
For the ``\emph{low}'' state, instead, the estimated value of $P_{\rm p}\sim 10^{47}$
erg s$^{-1}$ is a factor 10 less than in Nov 27, with $L_\gamma$ being
a factor 4 weaker.
To summarise: the total excursion (from ``\emph{low}'' state to Dec 3, 2009)
of $P_{\rm p}$ is factor 20, not very different to the total amplitude 
of $L_\gamma$ (factor $\sim$30).
But restricting the period from Nov 6 to Dec 3 the jet power is 
quasi--constant, while $L_\gamma$ varied by a factor 10.

\subsection{Jet power vs disk luminosity}

The disk luminosity is $L_{\rm d}= 6.7 \times 10^{46}$ erg s$^{-1}$,
comparable to $P_{\rm r}$.
The jet power must be greater than $P_{\rm r}$ {\it and therefore greater than
$L_{\rm d}$.}
This is one of the most important outcomes of having followed 3C 454.3 
during its major $\gamma$--ray flare.

Tab. \ref{para} lists the value of the outflowing mass rate,
estimated through $P_{\rm p} = \Gamma \dot M_{\rm out} c^2$.
Since $\Gamma$ varies moderately (between 15 and 20) while
$P_{\rm p}$ varies by a factor 10, we have that also
$\dot M_{\rm out}$ varies by an order of magnitude considering
the entire time span, and only by a factor 2 in Nov -- Dec 2009.
In this period it is of the order of 1 solar mass per year.
We can compare this value with the accretion mass rate $\dot M_{\rm in}$, 
that we can derive through $L_{\rm d}=\eta\dot M_{\rm in} c^2$.
Assuming $\eta=0.08$, a disk luminosity $L_{\rm d}=6.75 \times 10^{46}$
erg s$^{-1}$ gives $\dot M_{\rm in}=14.8 M_\odot$ yr$^{-1}$, about a factor
10 greater than $\dot M_{\rm out}$.

If the disk emits at a constant level, with a constant $\dot M_{\rm in}$,
we are forced to conclude that the link between the accretion rate
and the jet power is not determined by $\dot M_{\rm in}$, or, rather,
that this is not the only important quantity in producing $P_{\rm jet}$.
For the ensemble of bright FSRQs detected by {\it Fermi}, in fact, 
the disk luminosity does correlate with $P_{\rm jet}$, even
taking out the effect of a common redshift dependence in the two quantities 
(see Fig. \ref{ldlg} and the discussion in Ghisellini et al. 2010).
Since the spin of the black hole of 3C 454.3 is constant, the likely
quantity that modulates $P_{\rm jet}$ in Nov--Dec 2009 is the value of the 
magnetic field in the vicinity of the black hole horizon.
If we assume that $P_{\rm B,0}=P_{\rm jet}$ at the 
Schwarzschild radius $R_{\rm S}$, the magnetic field must be
\begin{equation}
B_0\equiv B(R_{\rm S}) \, =\, 10^5\, P^{1/2}_{\rm jet,48}
\left({5\times 10^8M_\odot \over M} \right)^{1/2}  \, \, {\rm G}
\end{equation}
In turn, we can check if the magnetic energy density $B_0^2/(8\pi)$ can be
equated to $\rho c^2$, the mass energy density of the accreting matter
close to the Schwarzschild radius. We can write
\begin{eqnarray}
\dot M_{\rm in} &=& 4\pi R_{\rm S}^2\, {H\over R_{\rm S}} \, \rho V_{\rm R} \, \to\, 
\nonumber \\
\rho c^2 &=& {\dot M_{\rm in}c^2  \over 4\pi R_{\rm S}^2 (H/R_{\rm S}) V_{\rm R}}
\nonumber \\
&=&  { L_{\rm d} \over  4 \pi \eta R_{\rm S}^2 (H/R_{\rm S}) V_{\rm R}} \, \to
\nonumber \\
{\rho c^ 2 \over B_0^2 /(8\pi) } 
&=&  { L_{\rm d} \over  P_{\rm jet}} \, {1\over 4 \eta  (H/R_{\rm S}) \beta_{\rm R}}
\end{eqnarray}
where $V_{\rm R}$ is the radial infalling velocity and $2 H$ is the total height of the disk
at the Schwarzschild radius.
The above equation shows that if $\rho c^ 2 \sim B_0^2 /(8\pi)$ and 
$P_{\rm jet}/L_{\rm d} \sim 10$, then $\beta_R (H/R)\sim 0.3$, i.e. not
an unreasonable value.

\subsection{Is 3C 454.3 exceptional?}

We finally ask if 3C 454.3 is exceptional, or instead if there are
other blazars reaching comparable values of $L_\gamma$ and
of $L_\gamma/L_{\rm d}$.
To answer, we can compare the flaring state of 3C 454.3 with all blazars 
of known redshift of the first 3--months {\it Fermi}/LAT all sky survey (Abdo
et al. 2009c), analysed in Ghisellini et al. (2010), as done in the top panel of
Fig. \ref{ldlg}.
We alert the reader that in Ghisellini et al. (2010) we have considered
the {\it average} $\gamma$--ray flux.
Had we considered the peak flux values there would be a few FSRQs 
of $L_\gamma$ even greater than 3C 454.3 in high state (one example
is PKS 1502+102, Abdo et al. 2010a).
In this respect we can conclude that 3C 454.3 is not the most
extreme blazar: its unprecedented $\gamma$--ray flux is due
to its relative vicinity in comparison to other FSRQs.
The same occurs in the planes $P_{\rm r}$--$L_{\rm d}$ and
$P_{\rm jet}$--$L_{\rm d}$ (mid and bottom panels of Fig. \ref{ldlg}):
there are other sources whose 3--months averaged $P_{\rm r}$ and
$P_{\rm jet}$ are comparable to 3C 454.3.

\section{Conclusions}

The study of the strong $\gamma$--ray flare of 3C 454.3,
together with the good coverage at X--ray and optical--UV
frequencies provided by {\it Swift} allowed to investigate
several issues about the physics of the jet of this blazar.
In Tavecchio et al. (2010) we analysed the behaviour of the 
1.5 year {\it Fermi}/LAT light curve, finding episodes of
very rapid variations, with time scales between 3 and 6 hours
both during the rising and the decaying phases.
This implies a very compact emitting region, suggesting that the
dissipation zone is not too far from the black hole, and that cooling
times are shorter than the variability time scales.
In turn this suggests that the dissipation region lies within the
BLR, at about one thousand Schwarzschild radii.

We have revisited the estimate of the mass of the black hole 
of 3C 454.3, finding a smaller value than found by 
Gu, Cao and Jiang (2001). 
Since in our model the mass of the black hole sets the
scales of the disk+jet system, a smaller mass helps to have
a more compact emitting region, that can vary on shorter
time scales.

The optical, X--ray and $\gamma$--ray fluxes correlate. 
This supports one--zone models.
The $\gamma$--ray flux varies quadratically (or even more)
with the optical and X--ray fluxes. 
By modelling the optical to $\gamma$--ray SED with a one--zone
synchrotron+inverse Compton leptonic model we can explain 
this behaviour if the magnetic field
is slightly fainter when the overall jet luminosity is stronger.

The power that the jet spent to produce the peak $\gamma$--ray luminosity
is of the same order than the accretion disk luminosity.
%During the flare, the total jet power surely surpassed the accretion power.  
Although the jet power correlates with the accretion luminosity considering
the ensemble of bright FSRQs detected by {\it Fermi}, 3C 454.3 probably
varied its jet power while maintaining a constant accretion luminosity.
This implies that the modulation of the jet power may not be due to 
variations of the accretion rate, but is probably due to variations
of the magnetic field close to its black hole horizon.

\section*{Acknowledgements}
We thank the referee for useful criticism.
We acknowledge the use of public data from the Swift data archive.
This research has made use of data obtained from the High 
Energy Astrophysics Science Archive Research Center (HEASARC), 
provided by NASA's Goddard Space Flight Center through the Science Support
Center (SSC).
Use of public \emph{GALEX} data provided by the Multimission Archive at the Space Telescope
Science Institute (MAST) is also acknowledged.
%Support from the Italian Space Agency, under contract ASI--INAF is acknowledged; 
 This work was partially financed by a 2007 COFIN-- MiUR grant and
by ASI grant I/088/06/0.


\begin{thebibliography}{99}

\bibitem[]{} Abdo A.A., Ackermann M., Ajello M. et al., 2009a, ApJ, 699, 817 % Early Fermi 3C 454.3
\bibitem[]{} Abdo A.A., Ackermann M., Ajello M. et al., 2009b, ApJSS, 183, 46  % Bright list, with TS description
\bibitem[]{} Abdo A.A., Ackermann M., Ajello M. et al., 2009c, ApJ, 700, 597 % Three Months List
\bibitem[]{} Abdo A.A., Ackermann M., Ajello M. et al., 2010a, ApJ, 710, 810  % 1502+106
\bibitem[]{} Abdo A.A., Ackermann M., Ajello M. et al., 2010b, ApJ, 710, 1271 % Spettri
\bibitem[]{} Aharonian F., Akhperjanian A.G., Anton G. et al., 2009, A\&A, 502, 749   % TeV - X correlation in 2155
\bibitem[]{} Atwood W.B., Abdo A.A., Ackermann M. et al., 2009, ApJ, 697, 1071 % LAT
\bibitem[]{} Barthelmy S.D., Barbier, L.M., Cummings, J.R., et al., 2005, Space Sci. Rev. 120, 143
\bibitem[]{} Bonning E., Bailyn C., Buxton M., et al.\ 2009, The Astronomer's Telegram, 2332, 1 
\bibitem[]{} Burrows D.N., Hill J.E., Nousek J.A., et al., 2005, Space Sci. Rev. 120, 165
\bibitem[]{} Buxton M., Bailyn C., Bonning E. et al., 2009, The Astronomer's Telegram, 2181, 1 
\bibitem[]{} Cardelli J. A., Clayton G. C. \& Mathis J. S., 1989, ApJ, 345, 245 
\bibitem[]{} Celotti A. \& Ghisellini G., 2008, MNRAS, 385, 283
\bibitem[]{} Escande L. \& Tanaka, Y.~T., 2009, The Astronomer's Telegram, 2328, 1 
\bibitem[]{} Foschini L., Tagliaferri G., Ghisellini G., et al. 2010,  MNRAS
  accepted, arXiv:1004.4518  
\bibitem[]{} Fossati G., Buckley J.H., Bond I.H. et al., 2008, ApJ, 677, 906 % TeV X correlation
\bibitem[]{} Frank J., King A. \& Raine D.J., 2002, Accretion power in astrophysics, Cambridge (UK) 
            (Cambridge University Press)
\bibitem[]{} Fuhrmann L., Cucchiara A., Marchili N. et al., 2006, A\&A, 445, L1
\bibitem[]{} Gehrels N., Chincarini G., Giommi P. et al., 2004, ApJ 611, 1005
\bibitem[]{} Georganopoulos M., Kirk J.~G., Mastichiadis A., 2001, ApJ, 561, 111
\bibitem[]{} Ghisellini G., Foschini L., Tavecchio F. \& Pian E., 2007, MNRAS, 382, L82
\bibitem[]{} Ghisellini G., Tavecchio F., Foschini L., Ghirlanda G., Maraschi L. \& Celotti A., 2010,  MNRAS, 402, 497 % General Properties 3-month Fermi
\bibitem[]{} Giommi P., Blustin A.J., Capalbi M. et al. 2006, A\&A, 456, 911  
\bibitem[]{} Gu M., Cao X. \& Jiang D.R., 2001, MNRAS, 327, 1111
\bibitem[]{} Gurwell M.A.,  2009, The Astronomer's Telegram, 2150, 1 
\bibitem[]{} Hill A.B., 2009, The Astronomer's Telegram, 2200, 1 
\bibitem[]{} Jackson N. \& Browne I.W.A., 1991, MNRAS, 250, 414 
\bibitem[]{} Jorstad S.G., Marscher A.P., Lister M. et al., 2005, AJ, 130, 1418 % big paper on VLBI superl.
\bibitem[]{} Jorstad S.G., Marscher A.P., Larionov V.M. et al., 2010, ApJ, 715, 362
\bibitem[]{} Kaspi S., Brandt W.N., Maoz D., Netzer H., Schneider D.P., \& Shemmer O.,
             2007, ApJ, 659, 997 
\bibitem[]{} Kaspi S., Smith P.S., Netzer H., Maoz D., Jannuzi B.T., \& Giveon U.,
             2000, ApJ, 533, 631 
\bibitem[]{} Katarzynski K. \& Ghisellini G., 2007, A\&A, 463, 529 % economico
\bibitem[]{} Katarzynski K. \& Walczewska K., 2010, A\&A, 510, A63 % variab in TeV blazars
\bibitem[]{} Katarzynski K., Ghisellini G., Tavecchio F., Maraschi L., Fossati G. 
             \& Mastichiadis A., 2005, A\&A, 433, 479 % X-TeV variab. in high-energy peaked BL Lac objects
\bibitem[]{} Krimm H.A., Barthelmy S.D., Baumgartner W. et al., 2009, The Astronomer's Telegram, 2330, 1 
\bibitem[]{} Lister M.L. \& Homan D.C., 2005, AJ, 130, 1389
\bibitem[]{} Lister M.L., Homan D.C., Kadler M. et al., 2009, ApJ, 696, L22 % superlum of Fermi
\bibitem[]{} Martin D.C., Fanson J., Schiminovich D. et al., 2005, ApJ, 619, L1  %GALEX MISSION OVERVIEW
\bibitem[]{} Mattox J.R., Bertsch D.L., Chiang J. et al., 1996, ApJ, 461, 396 
\bibitem[]{} McLure R.J. \& Dunlop J.S., 2001, MNRAS, 327, 199
\bibitem[]{} Moderski R., Sikora M., Coppi P.~S., Aharonian F., 2005, MNRAS, 363, 954 
\bibitem[]{} Morrissey P., Conrow T., Barlow T.A. et al. 2007, ApJSS, 173, 682 %GALEX DATA PRODUCTS
\bibitem[]{} Nandikotkur G., Jahoda K.M., Hartman R.C. et al., 2007, ApJ, 657, 706
\bibitem[]{} Netzer H., Brotherton M.S., Wills B.J., Han M.S., Wills D., Baldwin J.A., Ferland G.J., Browne I.W.A., 1995, ApJ, 448, 27
\bibitem[]{} Pian E., Falomo R., \& Treves A., 2005, MNRAS, 361, 919
\bibitem[]{} Pian E., Foschini L., Beckmann V., et al., 2006, A\&A, 449, L21 % 454.3 by Integral
\bibitem[]{} Pian E., Romani P., Treves A., et al., 2007, ApJ, 664, 106 % 0537 
\bibitem[]{} Poole T.S., Breeveld A.A., Page M.J. et al., 2008, MNRAS, 383, 627 
\bibitem[]{} Raiteri, C.~M., Villata M., Chen W.P. et al., 2008, A\&A, 485, L17
\bibitem[]{} Raiteri C.M., Villata M., Larionov V.M. et al., 2007, A\&A, 473, 819
\bibitem[]{} Rando R., 2009, arXiv:0907.0626 
\bibitem[]{} Roming P.W.A., Kennedy T.E., Mason K.O. et al., 2005, Space Sci. Rev. 120, 95
\bibitem[]{} Sakamoto T., D'Ammando F., Gehrels N., Kovalev Y.Y., \& Sokolovsky K., 2009, The Astronomer's Telegram, 2329, 1 
\bibitem[]{} Sikora M., \& Madejski G., 2000, ApJ, 534, 109
\bibitem[]{} Striani E., Vercellone S., Verrecchia F. et al., 2009a, The Astronomer's Telegram, 2322, 1 
\bibitem[]{} Striani E., Vercellone S., Verrecchia F. et al., 2009b, The Astronomer's Telegram, 2326, 1 
\bibitem[]{} Tavecchio F., Maraschi L., Ghisellini G., et al. 2002, ApJ, 575, 137
\bibitem[]{} Tavecchio F., Ghisellini G.,  Bonnoli G., Ghirlanda G., 2010, MNRAS, submitted
\bibitem[]{} Tosti G., Chiang, J., Lott, B., Do Couto E Silva, E., Grove, J.~E., \& Thayer, J.~G.\ 2008, The Astronomer's Telegram, 1628, 1 
\bibitem[]{} Vercellone S., Chen A.W., Giuliani A., et al., 2007, ApJ, 676, L13
\bibitem[]{} Vercellone S., D'Ammando F., Vittorini V. et al., 2010, ApJ, 712, 405
\bibitem[]{} Villata M., Raiteri C.M., Aller M.F. et al.,  2007, A\&A, 464, L5
\bibitem[]{} Villata M., Raiteri C.M., Balonek T.J. et al., 2006, A\&A, 453, 817 
\bibitem[]{} Villata M., Raiteri, C.M., Larionov, V.M., Konstantinova, T.S., Nilsson K., 
             Pasanen M., \& Carosati D., 2009, The Astronomer's Telegram, 2325, 1 
\bibitem[]{} Vlahakis N. \& K\"onigl A., 2004, ApJ, 605, 656 % accelerating jets
\bibitem[]{} Wills B.J., Thompson K.L., Han M. et al., 1995, ApJ, 447, 139 



\end{thebibliography}
\end{document}